\documentclass[journal]{IEEEtran}

\usepackage{amsmath,amsfonts, amssymb,epsfig,epstopdf,bm}
\usepackage{caption,lipsum,rotating,makecell,stfloats, eufrak, setspace,cite}
\usepackage{dsfont}
\usepackage{graphicx,multirow,multicol,threeparttable}
\graphicspath{{./}{Figures/}}
\usepackage{diagbox}
\usepackage{eufrak}
\usepackage[mathscr]{eucal}
\usepackage[table]{xcolor}
\usepackage{color}
\definecolor{BLUE}{rgb}{0,0,1}
\usepackage{colortbl}
\definecolor{mygray}{gray}{0.8}
\usepackage{url}
\usepackage{hyperref}
\usepackage{algorithm,algorithmic}
\newcommand{\tabincell}[2]{\begin{tabular}{@{}#1@{}}#2\end{tabular}}

\begin{document}

\title{Secure Reversible Data Hiding in Encrypted Images Using Cipher-Feedback Secret Sharing}

\author{{Zhongyun~Hua,~\IEEEmembership{Member,~IEEE,}
        Yanxiang~Wang,
        Shuang~Yi,
        Yicong~Zhou,~\IEEEmembership{Senior Member,~IEEE,}
        Xiaohua~Jia,~\IEEEmembership{Fellow,~IEEE,}}
\thanks{This work was supported in part by the National Natural Science Foundation of China under Grants 62071142 and 62002301, and the Guangdong Basic and Applied Basic Research Foundation under Grant 2021A1515011406, and the Natural Science Foundation of Chongqing under Grant cstc2019jcyj-msxmX0393, and the Education Committee foundation of Chongqing under Grant KJQN201900305, and the Research Committee at University of Macau under Grant MYRG2018-00136-FST.}
\thanks{Zhongyun~Hua, Yanxiang~Wang and Xiaohua~Jia are with School of Computer Science and Technology, Harbin Institute of Technology, Shenzhen, Shenzhen 518055, China (e-mail: huazyum@gmail.com).}
\thanks{Shuang Yi is with Engineering Research Center of Forensic Science, Chongqing Education Committee, College of Criminal Investigation, Southwest University of Political Science and Law, Chongqing 401120, China.}
\thanks{Yicong~Zhou is with Department of Computer and Information Science, University of Macau, Macau, China.}

}

\markboth{A manuscript submitted to IEEE Transactions,~2021}
{Shell \MakeLowercase{\textit{et al.}}: Bare Demo of IEEEtran.cls for Journals}

\maketitle

\begin{abstract}
Reversible data hiding in encrypted images (RDH-EI) has attracted increasing attention, since it can protect the privacy of original images while the embedded data can be exactly extracted. Recently, some RDH-EI schemes with multiple data hiders have been proposed using secret sharing technique. However, these schemes protect the contents of the original images with lightweight security level. In this paper, we propose a high-security RDH-EI scheme with multiple data hiders. First, we introduce a cipher-feedback secret sharing (CFSS) technique. It follows the cryptography standards {\color{black}by introducing the cipher-feedback strategy of AES}. Then, using the CFSS technique, we devise a new $(r,n)$-threshold ($r\leq n$) RDH-EI scheme with multiple data hiders {\color{black}called CFSS-RDHEI}. It can encrypt an original image {\color{black}into $n$ encrypted images} with reduced size {\color{black}using an encryption key} and sends each {\color{black}encrypted image} to one data hider. Each data hider can independently embed secret data into the {\color{black}encrypted image} to obtain the corresponding marked encrypted image. The original image can be completely recovered from $r$ marked encrypted images {\color{black}and the encryption key}. Performance evaluations show that {\color{black}our CFSS-RDHEI} scheme has high embedding {\color{black}rate} and its generated {\color{black}encrypted images} are much smaller, compared to {\color{black}existing} secret sharing-based RDH-EI schemes. Security analysis demonstrates that it can achieve high security to defense some commonly used security attacks.

\end{abstract}


\section{Introduction}
\label{Section1}
With the fast development of cloud computing, there is an increasing need of privacy protection. To keep the data privacy of holders while to perform necessary operations, multimedia signal processing in encrypted domain has attracted increasing attention~\cite{zhang2014compressing}. To facilitate the data management of encrypted image in cloud environment, reversible data hiding in encrypted image (RDH-EI) technique has been developed~\cite{puech2008reversible,zhang2011reversible}. The RDH-EI contains three participants: content owner, data hider and receiver. The content owner encrypts the original images to protect the contents. The data hider can embed some secret data into the encrypted images without accessing the contents of original image. Subsequently, the receiver can extract the embedded data and recover the original image losslessly.

In 2008, Puech $et~al.$ first proposed a RDH-EI scheme~\cite{puech2008reversible}, in which one bit can be embedded into an image block of cipher-image encrypted by the advanced encryption standard (AES), and the embedded bit can be extracted by analyzing the local standard deviation of the image block. Several years later, Zhang proposed another new RDH-EI scheme~\cite{zhang2011reversible}. Inspired by these works, many new RDH-EI schemes have been developed using different techniques~\cite{liao2015reversible, qin2015effective, zhang2011separable}. According to the data embedding strategy, all the RDH-EI schemes can be roughly classified as two types: vacating room after encryption (VRAE)~\cite{zhang2011reversible,liao2015reversible,qin2015effective,zhang2011separable} and reserving room before encryption (RRBE)~\cite{ma2013reversible, zhang2014reversibility, zhang2015lossless, yi2017binary, puteaux2018efficient}. The VRAE strategy first encrypts an original image and then embeds secret data into the encrypted image, while the vacated room for data embedding in the RRBE strategy is pre-reserved before encryption. Compared with the VRAE strategy, the RRBE strategy can take advantage of the spatial correlation of the original image and usually provides a larger embedding capacity than the VRAE strategy. Some typical examples of the RRBE strategy are as follows. Puteaux and Puech proposed an RDH-EI scheme by predicting the most significant bit (MSB) of the pixels~\cite{puteaux2018efficient}. This scheme can make fully use of the local correlation of the plain-image and achieves a high embedding capacity. Using this strategy, Yi $et~al.$ further proposed a two-MSBs prediction method~\cite{puyang2018reversible} and Yin $et~al.$~\cite{yin2019reversible} designed a variable length of MSB prediction method using Huffman coding.

To keep redundant information for data embedding, existing RDH-EI schemes using VRAE or RRBE strategy usually encrypt the original images using some lightweight encryption methods such an xor operation~\cite{yi2017binary,tuncer2016reversible,avci2016novel} and block-based permutation~\cite{aryal2017integrated, qin2019efficient}. Although these operations can achieve noise-like cipher-images, it is obvious that the encrypted results cannot defense many security attacks such as the differential attack and chosen-plaintext attack. To enhance the security, some new RDH-EI schemes have been developed using the homomorphic encryption~\cite{wu2019homo1, xiong2019homo2, jiang2020homo3, zheng2019homo4}. In these schemes, the original images are encrypted {\color{black}using} some homomorphic encryption algorithms and {\color{black}the secret data are} embedded into the encrypted images {\color{black}using} the homomorphic properties. However, since all the homomorphic encryption algorithms have extremely high computation cost and data expansion, the encryption efficiency is greatly reduced and the sizes of encrypted images are expanded.

Recently, some RDH-EI schemes have been proposed using secret sharing technique~\cite{wu2018adopting,chen2019newsecret,chen2020secret}. \textcolor{black}{In~\cite{wu2018adopting}, Wu $et~al.$ first adopted secret sharing technique for RDH-EI scheme. The original images are first encrypted using Shamir's secret sharing scheme and the secret data are then embedded into the encrypted images using difference expansion or difference histogram shifting.} Only with sufficient number of image shares, one can recover the original image. \textcolor{black}{Later, Chen $et~al.$~\cite{chen2019newsecret} proposed another secret sharing-based RDH-EI method. In this work, the data hider embeds secret data using the difference expansion and the addition homomorphism. However, this work has only one data hider and the embedding rate isn't high. To achieve a larger embedding rate, Chen $et~al.$~\cite{chen2020secret} proposed a new method with multiple data hiders. In the data hiding phase, each data hider can embed secret data by substituting one pixel in every $n$ pixels. The embedding rate in this method is reduced when the number of data hiders increases and it is not applicable when the number of data hider is more than seven.}

In this paper, we propose a new RDH-EI scheme using a novel secret sharing method. It can achieve a high security while {\color{black}keeping} a large embedding rate. First, we present a novel secret sharing technique called cipher-feedback secret sharing (CFSS). It follows the cryptography standards. Using the CFSS, we further propose a secure RDH-EI scheme with multiple data hiders. In an $(r,n)$-threshold ($r\leq n$) scheme, the proposed RDH-EI can encrypt an original image {\color{black}into $n$ encrypted images} and sends each {\color{black}encrypted image} to one data hider. Since $r-1$ pixels are processed in one-time sharing, each {\color{black}encrypted image} is the $1/(r-1)$ size of the original image. Each data hider can independently embed secret data into the {\color{black}encrypted image} to obtain the corresponding marked encrypted image. Since the used error flag strategy~\cite{puteaux2018efficient,puyang2018reversible} in most existing MSB prediction can cause security problems~\cite{dragoi2020security}, a multi-MSBs prediction method {\color{black}with secure location map} is used to embed secret data.
{\color{black}The main contributions and novelty of this paper are summarized as follows:

\begin{enumerate}
    \item 	We develop a cipher-feedback secret sharing (CFSS) technique applying the cipher-feedback strategy of AES to the polynomial-based secret sharing scheme. The CFSS follows the cryptography standards and can share images with higher security than existing secret sharing techniques;

    \item Using the CFSS, we propose an $(r,n)$-threshold $(r \leq n)$ RDH-EI scheme called CFSS-RDHEI. Different from other secret sharing-based schemes~\cite{wu2018adopting,chen2019newsecret,chen2020secret} that share image pixels independently, our CFSS-RDHEI shares pixels using the cipher-feedback strategy and its generated encrypted images have higher security and smaller size than these produced by schemes in~\cite{wu2018adopting,chen2019newsecret,chen2020secret};

    \item A multi-MSBs prediction method with a secure location map is used to embed secret data. It can avoid the security problems~\cite{dragoi2020security} caused by the error flag in most existing MSB prediction methods and achieves larger embedding rate than these methods;

    \item Experiment results verify that our CFSS-RDHEI can achieve a much larger embedding rate and its encrypted images are much more secure and smaller, compared with existing secret sharing-based RDH-EI schemes.

\end{enumerate}
}

The rest of this paper is organized as follows. Section~\ref{Section2} introduces the developed CFSS. Section~\ref{Section3} presents the CFSS-RDHEI in detail. Section~\ref{Section4} simulates the CFSS-RDHEI and analyzes its embedding performance. Section~\ref{Section5} analyzes the security of the CFSS-RDHEI and Section~\ref{Section6} concludes this paper.

\section{Cipher-Feedback Secret Sharing}
\label{Section2}
In this section, we introduce the concept of the polynomial-based secret sharing technique, discuss its properties, and then develop a cipher-feedback secret sharing (CFSS) with a high security level.
\subsection{Existing Polynomial-Based Secret Sharing}

The $(r,n)$-threshold ($r\leq n$) polynomial-based secret sharing technique was first proposed by Shamir~\cite{shamir1979share}, where a dealer takes a secret as input and generates $n$ shares for $n$ parties (each party holds one share). A collection of any $r$ shares can recover the original secret. The polynomial-based secret sharing technique can be used to solve many research issues such as acting as a cryptographic scheme.

The fundamental theory of the polynomial-based secret sharing is to construct a group of polynomial equations over a finite filed $F$, which satisfies that $F>n$. According to some known theorems, any $r$ pairs $(x_i,y_i)\in F\times F$ with different $\{x_i\}$ can uniquely determine a polynomial $f$ of degree $(k-1)$ such that $f(x_i)=y_i$ for $i\in\{1,2,\cdots,k\}$.

The $(r,n)$-threshold secret sharing scheme to a secret $S$ can be constructed as follows. First, let $\{x_1,x_2,\cdots,x_n\}$ be $n$ different random positive integers that are all smaller than $F$. Then an $r-1$ degree polynomial can be constructed by
\begin{equation}
\label{eqa.shamirpoly}
f(x)= (a_0 + \sum_{k=1}^{r-1}a_k x^{k} )\mod \ F,
\end{equation}
where the $r$ coefficients $a_{0},\cdots,a_{r-1}$ can be any random integers or the elements of the secret $S$ and they are all smaller than $F$ as well. A party $P_i$ holds $(x_i,f(x_i))$, where $x_i$ is the identity and $f(x_i)$ is the share.

When any $r$ shares with their identities have been collected (e.g. $(x_i,f(x_i)), i=1,2,\cdots,r$), one can reconstruct the polynomial $f$ using the Lagrange interpolation as
\begin{equation}
\label{eqa.lagr}
L(x)=\sum_{j=1}^{r}(f(x_j)\times \prod_{\substack{0 < k \leq r\\k\neq j}}\frac{x - x_k}{x_j - x_k}) \mod\ F.
\end{equation}
Then the first coefficient $a_0$ can be obtained as $a_0=L(0)$. Meanwhile, a completely expanded $L(x)$ can reveal the other coefficients $a_{1},\cdots,a_{r-1}$ of the original polynomial as well. Then the secret $S$ can be obtained.

In the original Shamir's secret sharing scheme~\cite{shamir1979share}, only the constant item $a_0$ is taken from the secret and the other $r-1$ coefficients $a_{1},\cdots,a_{r-1}$ are all randomly selected integers. In this scheme, the $n$ generated shares have the same size with the original secret. Since all $r$ coefficients $a_{0},\cdots,a_{r-1}$ can be correctly recovered, part or all of them can be the elements of the secret. In the Thien-Lin secret sharing scheme~\cite{thien2002secret}, all the $r$ coefficients are taken from the secret and thus the shares have a smaller size than the original secret.

\subsection{CFSS}
Because of the sharing strategy, the polynomial-based secret sharing technique has been widely used as a cryptographic scheme in many scenarios. However, this scheme cannot achieve diffusion property and thus has weak ability to resist many security attacks such as the differential attack and chosen-plaintext attack. The diffusion property indicates that a bit change in the plaintexts should cause the half difference in the ciphertexts. Without diffusion property, one can recover the {\color{black}sharing parameters} $x_1,x_2,\cdots,x_n$ using the differential attack or chosen-plaintext attack. Thus, the existing polynomial-based secret sharing has security drawbacks when measured by strict cryptography standards.

\begin{figure}[htbp]
	\centering
	\begin{minipage}[b]{0.8\linewidth}
		\centerline{\includegraphics[width=1\linewidth]{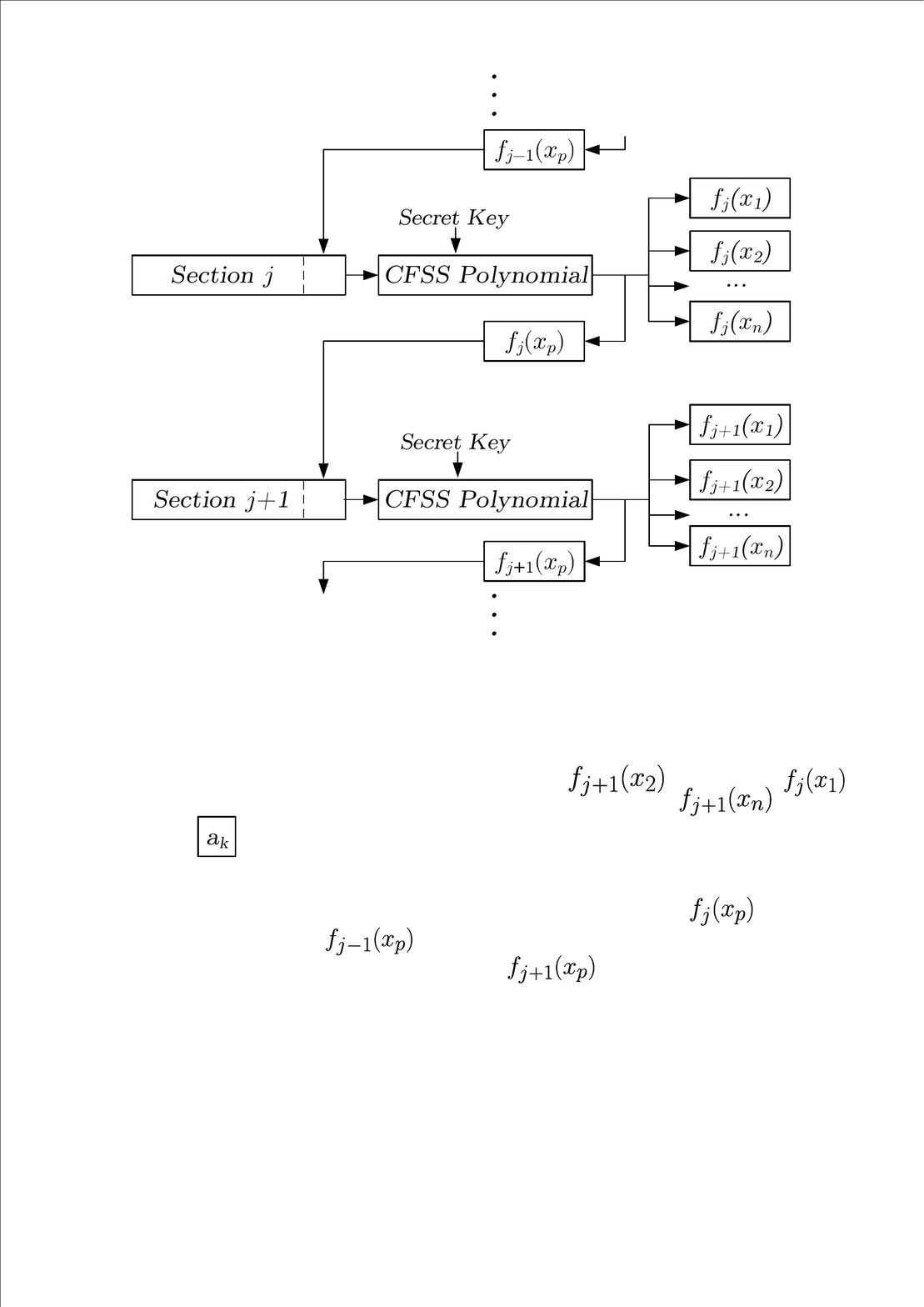}}
	\end{minipage}
	\caption{The demonstration of the developed CFSS.}
	\label{fig.cfss}
\end{figure}

To address the security issues of existing secret sharing, we introduce a new secret sharing method called CFSS {\color{black}using} the cipher-feedback mode of AES. Fig.~\ref{fig.cfss} shows the main concept of the CFSS. It can be seen that the last coefficient of the polynomial in the current sharing operation is taken from a random share (i.e., $p$ is randomly selected among $1,2,\cdots,n$ in each sharing) of the previous sharing result. The last coefficient in the first sharing operation is a random integer. Using this cipher-feedback strategy, any change in the secret image can cause complete difference in the obtained shares.

Then the $(r,n)$-threshold CFSS can be constructed as follows. (1) Divide the secret $S$ into several non-repeated sections and each section has $r-1$ elements; (2) Generate $n$ different integers $\{x_1,x_2,\cdots,x_n\}$ from the secret key; (3) For the $j$-th section, an $(r-1)$ degree polynomial is constructed by

\begin{equation}
\label{eqa.share}
f_j(x)= (a_0+ \sum_{k=1}^{r-2}a_k x^{k} + f_{j-1}(x_p)x^{r-1} )\mod \ F ,
\end{equation}
where $a_0,\cdots,a_{r-2}$ are $r-1$ elements from the $j$-th section of $S$ and $f_{j-1}(x_p)$ is a randomly selected previous sharing result. For the first section $j=1$, $f_0(x_p)$ {\color{black}can be any} random integer satisfying that $f_0(x_p) < F$. {\color{black}When} $x\in\{x_1,x_2,\cdots,x_n\}$, $n$ shares can be generated as $\{f_j(x_1),f_j(x_2),\cdots,f_j(x_n)\}$. Then a party $P_i$ holds the {\color{black}identity $i$ and $f_j(x_i)$, which is the $j$-th element of the share}.

The $(r,n)$-threshold CFSS scheme can achieve the following properties. (1) It can achieve diffusion property. When a bit is changed in the secret $S$, all the elements can be randomly changed by the cipher-feedback structure and the random integer $f_0(x_p)$. {\color{black}Thus, the CFSS can achieve high security. It follows the strictly cryptography standards and generates shares with high ability to resist some commonly used security attacks.} (2) The security level can be further improved, because $f_{j-1}(x_p)$ is randomly selected from previous $n$ sharing results, and $p$ can be different in each sharing. (3) Since each section has $(r-1)$ elements and generates one element for each share, the size of the obtained share can be greatly reduced to $1/(r-1)$ size of the original secret $S$.

\section{CFSS-Based RDH-EI}
\label{Section3}
\begin{figure*}[htbp]
\centering
\begin{minipage}[b]{0.9\linewidth}
  \centerline{\includegraphics[width=1\linewidth]{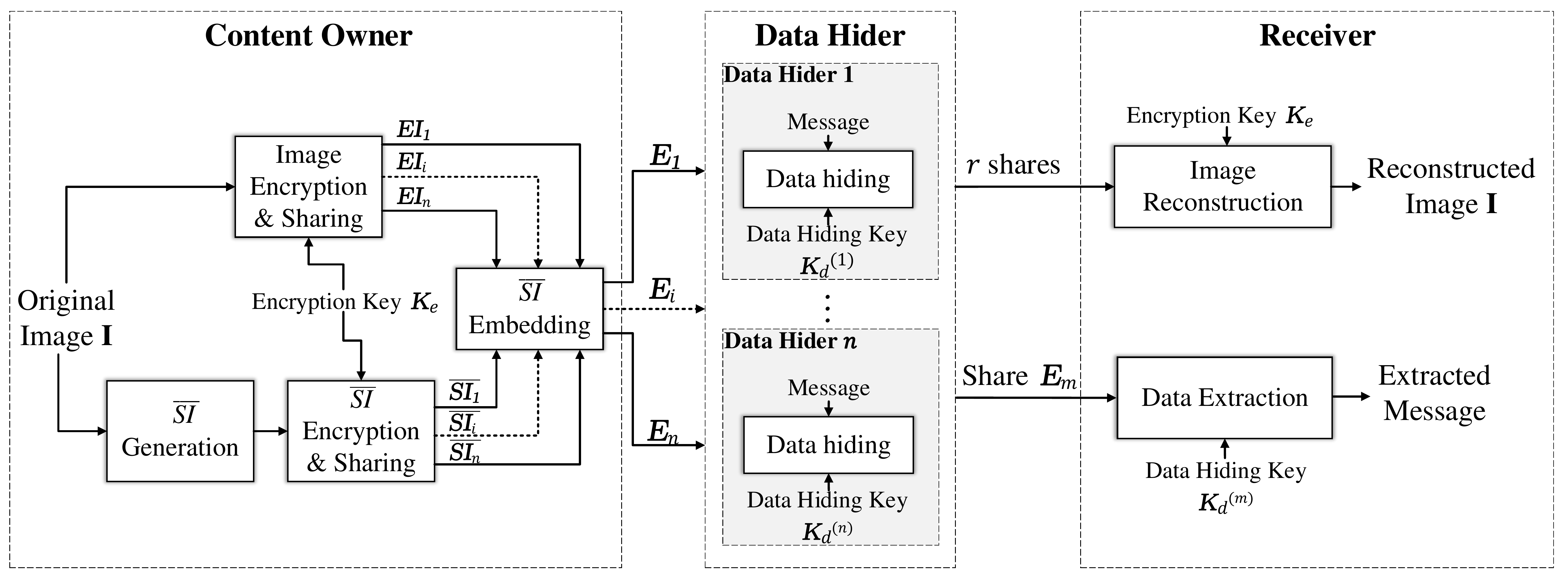}}
    \end{minipage}
\caption{{\color{black}The overview of the CFSS-RDHEI.}}
\label{fig.struct}
\end{figure*}

In this section, we propose a new RDH-EI using the CFSS called CFSS-RDHEI. Fig.~\ref{fig.struct} shows the structure of the CFSS-RDHEI. The MSB prediction can make full use of the high adjacent pixel correlations of a natural image. However, according to the discussions in~\cite{dragoi2020security}, the error flag strategy in {\color{black}most existing MSB prediction methods may} cause security problems. To address this issue, \textcolor{black}{a multi-MSBs prediction method with location map is used to embed secret data. The location map is first compressed and then encrypted by the CFSS into $n$ shares, which can ensure a security level. We call the image shares before data embedding as the encrypted images and after data embedding as the marked encrypted images. }

\textcolor{black}{In the CFSS-RDHEI, the initialization process first calculates the proportion of precisely predictable pixel number $P_c$, and then determines the optimal level $l$ using $P_c$. Then the final side information is calculated. After that, the final side information and the $(8-l)$ LSBs of the original image are encrypted into $n$ shares using the CFSS scheme with an encryption key. After embedding $n$ shares of side information into $n$ encrypted images, the encrypted images are sent to data hiders.} Once a data hider receives \textcolor{black}{an encrypted image}, he/she can embed additional data into \textcolor{black}{the encrypted image} by substituting the multi-MSBs of available pixels {\color{black}to obtain the marked encrypted images}. At the receiver side, one owning at least $r$ \textcolor{black}{marked encrypted images and the encryption} key can reconstruct the original image losslessly.


\subsection{Side Information Generation}
\label{sec.si}
The CFSS-RDHEI uses the multi-MSBs substitution {\color{black}with secure location map} to embed the secret data. The multi MSBs of pixels should be correctly predicted when reconstructing the original image. However, some pixels with special values cannot be correctly predicted. Their values and positions should be stored.

The median edge detector (MED) predictor is a high-performance predictor~\cite{weinberger2000loco} and it is used in our method to predict the multi-MSBs of image pixels. Using the previous adjacent pixels, the MED predictor can predict the current pixel value. 
The expression of the MED predictor used in the proposed method is shown as follows,

\begin{equation}
\label{equ.med}
\tilde{x}_{i,j}=
\begin{cases}
b, & \quad \mbox{for}\ j = 1, i\neq 1; \\
c, & \quad \mbox{for}\ i = 1, j \neq 1; \\
\max(b,c),& \quad \mbox{for}\ a \leq \min(b,c);\\
\min(b,c),& \quad \mbox{for}\ a \geq \max(b,c);\\
b + c - a,& \quad \mbox{otherwise},
\end{cases}
\end{equation}
where $a, b$, and $c$ are the upper-left, upper, and left adjacent pixels of $x_{i,j}$, respectively. Note that the first pixel $x_{1,1}$ isn't processed.

Since different images may have greatly different smoothness, their optimal embedding capacity can be achieved at different levels of MSBs. To achieve a large embedding {\color{black}rate} for each image, an optimal level $l$ for $l$-MSBs prediction should be first determined. For simplicity, {\color{black}the proportion of pixels $P_c$ that can be precisely predicted} is used to detect the $l$. For an image of size $M\times N$, the $P_c$ is calculated as
{\color{black}
\begin{equation}
\label{eqa.pc}
P_c = \frac{\sum\limits_{(i,j\in[1,M]\times[1,N])\cap (i,j)\neq(1,1)}P(i,j)}{M\times N}.
\end{equation}}
where $P(i,j)$ is the prediction correctness of each pixel and it is defined as
\begin{equation}
\label{eqa.pc2}
P(i,j)=\begin{cases}
1 & \quad \mbox{for}\ x_{i,j} = \tilde{x}_{i,j};\\
0 & \quad \mbox{otherwise}.
\end{cases}
\end{equation}

\textcolor{black}{According to the experience-based learning on a large number of natural images, the $l$ is set as}
\textcolor{black}{
\begin{equation}
\label{equ.thres}
l=
\begin{cases}
4 & \quad \mbox{for}\ P_c \leq 0.063 ; \\
5 & \quad \mbox{for}\ 0.063 < P_c \leq 0.102 ; \\
6 & \quad \mbox{for}\ P_c > 0.102.
\end{cases}
\end{equation}}

Denote $x_{i,j}^{lMSB}$ and $\tilde{x}_{i,j}^{lMSB}$ as the \textit{l}-MSBs values of the original and predicted pixel, respectively. Their values can be calculated as
{\color{black}
\begin{equation}
\label{eqa.3msb}
\begin{cases}
x_{i,j}^{lMSB}=(x_{i,j}-(x_{i,j}\mod\ 2^{8-l}))\times 2^{-(8-l)} ,\\
\tilde{x}_{i,j}^{lMSB}=(\tilde{x}_{i,j}-(\tilde{x}_{i,j}\mod\ 2^{8-l}))\times 2^{-(8-l)}.
\end{cases}
\end{equation}}
The $l$-MSBs prediction error between $x_{i,j}^{lMSB}$ and $\tilde{x}_{i,j}^{lMSB}$ is calculated as

\begin{equation}
\label{eqa.pe}
Lpe = x_{i,j}^{lMSB} - \tilde{x}_{i,j}^{lMSB}.
\end{equation}

\begin{figure}[!htbp]
	\centering
	\begin{minipage}[b]{0.99\linewidth}
		\begin{minipage}[b]{0.49\linewidth}
			\centerline{\includegraphics[width=1\linewidth]{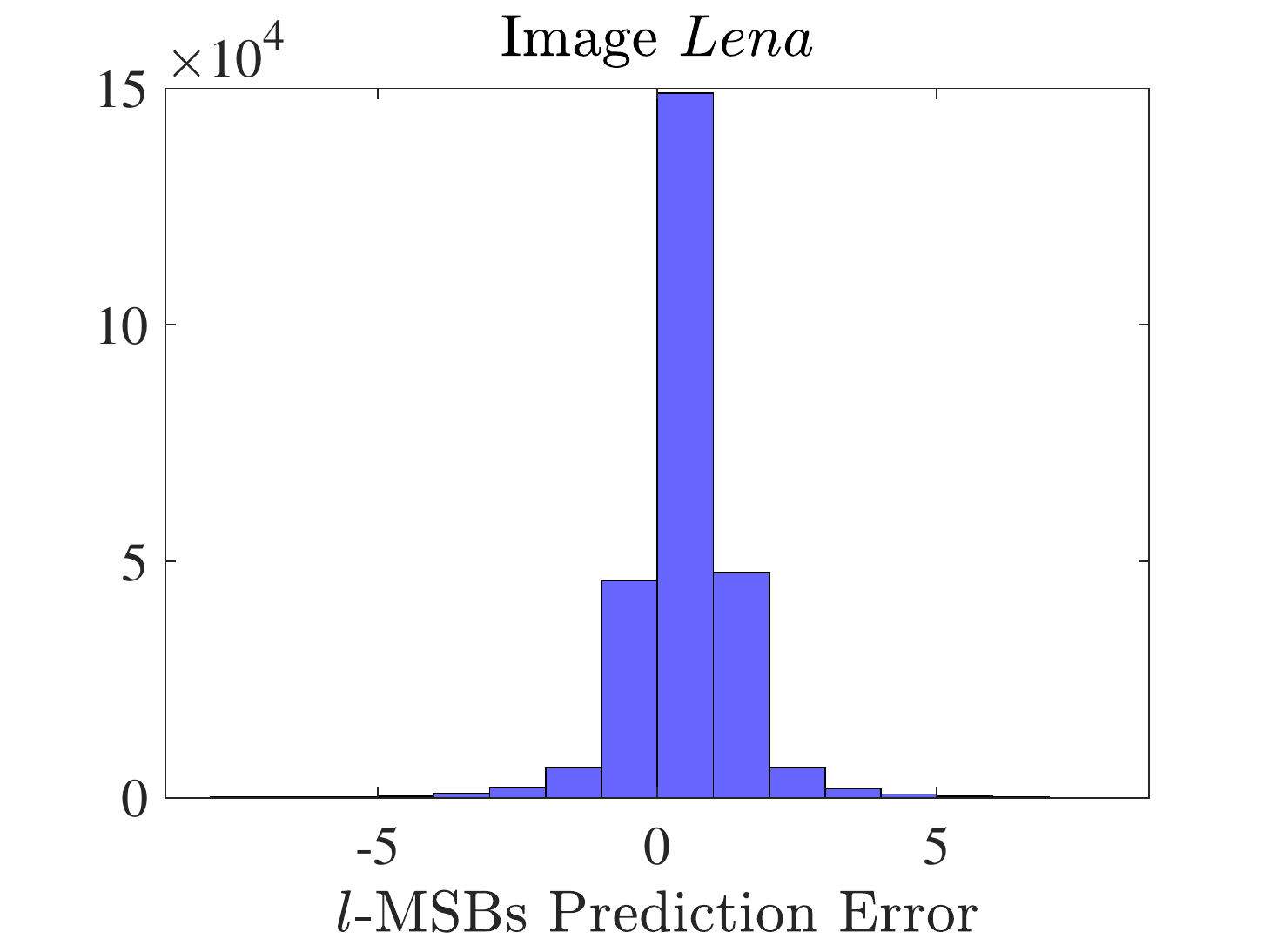}}
			\centerline{(a)}
		\end{minipage}\hfill
		\begin{minipage}[b]{0.49\linewidth}
			\centerline{\includegraphics[width=1\linewidth]{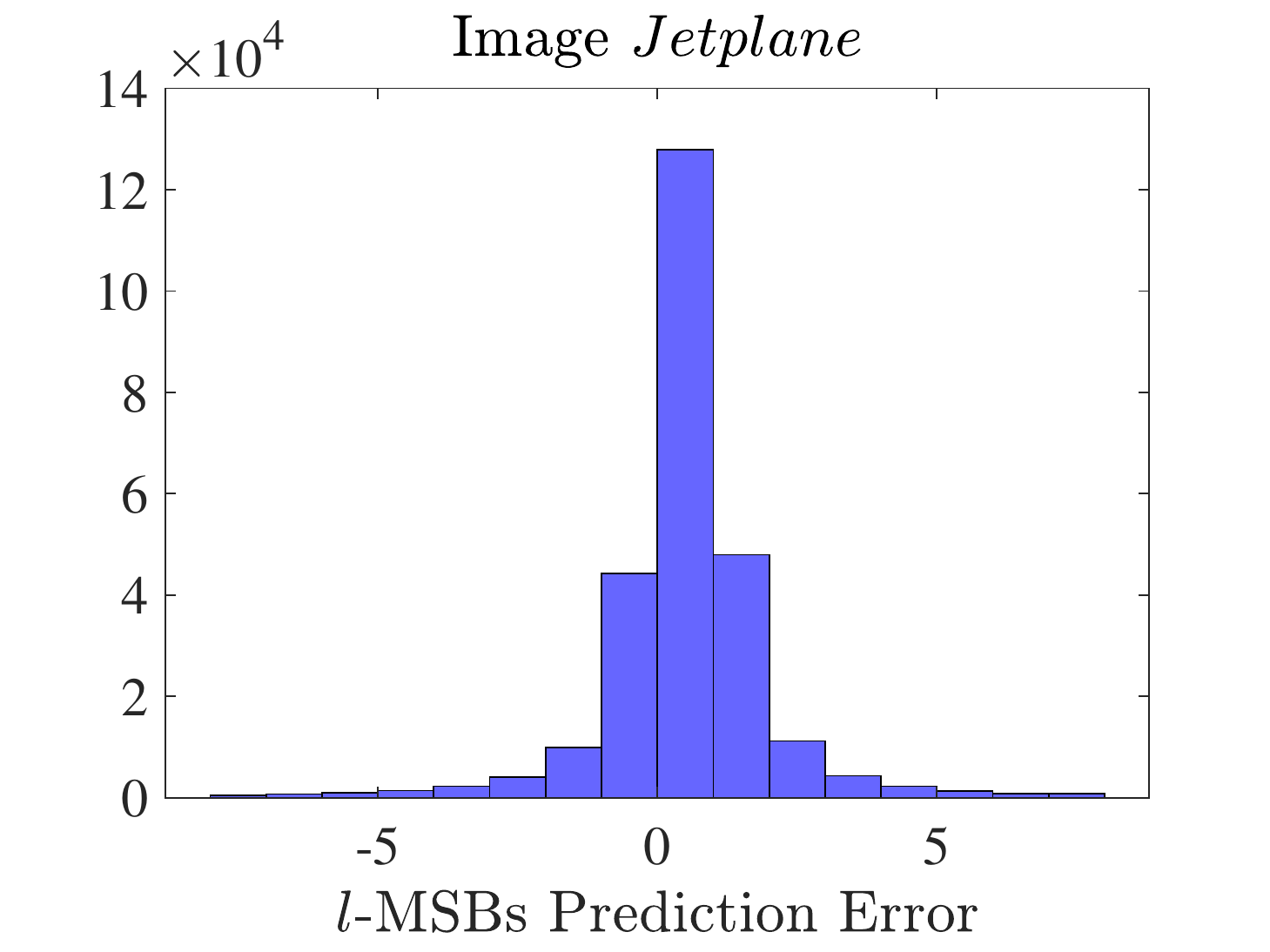}}
			\centerline{(b)}
		\end{minipage}\hfill\\
	\end{minipage}\hfill
	\caption{The histograms of the $l$-MSBs prediction errors in different test images. (a) Image \textit{Lena} with $l=5$; (b) image \textit{Jetplane} with $l=6$.}
	\label{fig.pehisto}
\end{figure}

\textcolor{black}{Because a natural image has high adjacent pixel correlation, a pixel can be estimated using its adjacent pixels. Thus, if the $l$ MSBs of a pixel can be completely predicted, its $l$-MSBs predication error $Lpe$ is 0. If the difference between $x_{i,j}^{lMSB}$ and $\tilde{x}_{i,j}^{lMSB}$ of a pixel is 1, its $l$-MSBs predication error $Lpe$ is -1 or 1. Similarity, if the difference between $x_{i,j}^{lMSB}$ and $\tilde{x}_{i,j}^{lMSB}$ of a pixel is 2, its $l$-MSBs predication error $Lpe$ is -2 or 2. Fig.~\ref{fig.pehisto} displays the histograms of the $l$-MSBs prediction errors of two commonly used natural images. It can be seen that the numbers of pixels with $l$-MSBs predication errors $-1$, 0, and $1$ are the largest. This is because a natural image has high adjacent pixel correlation.}
To fully utilize this statistical characteristic, we propose a new method to store the extra information. First, a location map is used to mark the pixels. \textcolor{black}{A pixel is a predictable pixel if its prediction error $Lpe\in \{ -1, 0, 1\}$ and its related position on the location map is marked as 0; otherwise, it is an unpredictable pixel and its position on the location map is marked as 1.} Since most pixels are predictable pixels, the location map has high data redundancy (e.g., more than $ 90\% $ elements of the location map are 0 in Fig.~\ref{fig.pehisto}). After being compressed using the compression scheme introduced in~\cite{chen2019high}, a compressed location map \textbf{\textit{LM}} is generated as a part of the side information. Since there are three types of predicable pixels, namely the $l$-MSBs prediction error {\color{black}$Lpe$ is $-1$, 0, or $1$}, we encode {\color{black}these three types of prediction errors using three codes, $10$, $0$, and $11$, respectively. Then the prediction error codes} \textbf{\textit{Lpes}} can be generated by
{\color{black}
\begin{equation}
\label{eqa.lpe}
\textbf{\textit{Lpes}}=\begin{cases}
\textbf{\textit{Lpes}}\mathbin\Vert 0 & \quad \mbox{for}\ Lpe= 0; \\
\textbf{\textit{Lpes}}\mathbin\Vert 10 & \quad \mbox{for}\ Lpe= -1;\\
\textbf{\textit{Lpes}}\mathbin\Vert 11 & \quad \mbox{for}\ Lpe= 1.
\end{cases}
\end{equation}}
where '$\mathbin\Vert$' stands for concatenation symbol.

The $l$ MSBs of all the unpredictable pixels should be stored. \textcolor{black}{According to the median edge detector used in Eq.~\eqref{equ.med}, only the first pixel cannot be processed and its $l$ MSBs should also be stored. The $l$ MSBs of the first pixel and all the unpredictable pixels} are combined to obtain a binary sequence \textbf{\textit{B}}, which is another part of the side information. 

After processing all the $l$ MSBs of the image pixels, the $(8-l)$ LSBs of the image are encrypted using the proposed CFSS method. \textcolor{black}{To achieve a high efficiency, several adjacent pixels are concatenated to form a larger number and the modular coefficient $F$ in Eq.~\eqref{eqa.share} is set as the largest prime integer that is smaller than or equal to the largest concatenated element. Table~\ref{table.ImaCon} shows the pixel concatenation strategies. For example, if $l=4$, two adjacent pixels are concatenated and the largest concatenated value is $2^8-1$, namely 255. Then set $F=251$.}

\begin{table}[!htbp]
\small
	\renewcommand{\arraystretch}{1}
	\setlength{\tabcolsep}{4pt}
	\begin{center}
		\caption{{\color{black}Pixel concatenation strategies in the $(8-l)$-LSBs image for different optimal level $l$.}}
		\label{table.ImaCon}
		\begin{tabular}{>{\color{black}}c>{\color{black}}c>{\color{black}} c}
			\hline
			Optimal level $l$  & Concatenated pixel number & Largest prime $F$ \\
			\hline
             4 & 2 & 251 \\
             5 & 2 & 61 \\
             6 & 4 & 251 \\
			\hline
		\end{tabular}
	\end{center}
\end{table}

Then, {\color{black}for these concatenated elements $I^{c}_{i,j}$ whose values are bigger than or equal to $F$, we change their values to $F-1$.} The modified information should be stored to recover the original pixels. The reference information \textbf{\textit{T}} is to {\color{black}encode the difference between $F-1$ and the concatenated elements $I^{c}_{i,j}$ that is bigger than or equal to $F$}. For example, if $F=251$, the reference information \textbf{\textit{T}} can be generated as

\begin{equation}
\label{eqa.t}
\textbf{\textit{T}}=\begin{cases}
\textbf{\textit{T}}\mathbin\Vert 101 & \quad \mbox{for}\ I^{c}_{i,j}=255;\\
\textbf{\textit{T}}\mathbin\Vert 100 & \quad \mbox{for}\ I^{c}_{i,j}=254;\\
\textbf{\textit{T}}\mathbin\Vert 011 & \quad \mbox{for}\ I^{c}_{i,j}=253;\\
\textbf{\textit{T}}\mathbin\Vert 010 & \quad \mbox{for}\ I^{c}_{i,j}=252;\\
\textbf{\textit{T}}\mathbin\Vert 001 & \quad \mbox{for}\ I^{c}_{i,j}=251;\\
\textbf{\textit{T}}\mathbin\Vert 000 & \quad \mbox{for}\ I^{c}_{i,j}=250;\\
\textbf{\textit{T}} & \quad \mbox{otherwise}.
\end{cases}
\end{equation}
{\color{black}
Note that only two bits are required to encode the difference between $F-1$ and $I^{c}_{i,j}$ when $F=61$.

Then the side information \textbf{\textit{SI}} consists of the compressed location map \textbf{\textit{LM}}, $l$-MSBs prediction error codes \textbf{\textit{Lpes}}, $l$ MSBs of the unpredictable pixels \textbf{\textit{B}}, and the reference information \textbf{\textit{T}}, namely \textbf{\textit{SI}}= \textbf{\textit{LM}} $\mathbin\Vert$ \textbf{\textit{Lpes}} $\mathbin\Vert$ \textbf{\textit{B}} $\mathbin\Vert$ \textbf{\textit{T}}. For an image of size $M\times N$, the maximum lengths of \textbf{\textit{LM}}, \textbf{\textit{Lpes}} and \textbf{\textit{B}} are $MN$, $2MN$ and $lMN$, respectively. For the \textbf{\textit{T}}, when $l=4$ with $F=251$, it has the theoretical maximum length $3MN/2$. Thus, we use $\left\lceil \log_2(MN) \right\rceil$ bits, $\left\lceil \log_2(2MN) \right \rceil$ bits, $\left\lceil \log_2(lMN) \right \rceil$ bits and $\left \lceil \log_2(3MN/2) \right \rceil$ bits to record the lengths of \textbf{\textit{LM}}, \textbf{\textit{Lpes}}, \textbf{\textit{B}} and \textbf{\textit{T}}, respectively. Besides, since \textbf{\textit{LM}} is obtained by compressing the location map using the method in~\cite{chen2019high} and this method contains four possible rearrangement types, another 2 bits are used to record the rearrangement type. All these additional bits are put in the front of \textbf{\textit{SI}}. The final side information $\overline{\textbf{\textit{SI}}}$ consists of the \textbf{\textit{SI}} and the bit sequence \textbf{\textit{Tsi}} for pre-processing \textbf{\textit{SI}}, which is discussed in Section~\ref{Section.SIE}. After being encrypted using the CFSS, each share of $\overline{\textbf{\textit{SI}}}$ is embedded into the vacated rooms of each encrypted image.}

\subsection{$(r,n)$-Threshold Sharing and Encryption}
\label{sec.enc}

In the content owner side, the $(8-l)$ LSBs of the original image are encrypted into $n$ \textcolor{black}{encrypted images using CFSS with an encryption key. Thus, encryption in the CFSS-RDHEI scheme indicates the secret sharing process}. To losslessly recover the whole image $I$, the final side information $\overline{\textbf{\textit{SI}}}$ should be used {\color{black}in the recovering process} and it {\color{black}should be} embedded into the \textcolor{black}{encrypted images}. We also encrypt $\overline{\textbf{\textit{SI}}}$ into $n$ shares and then embed each share into one \textcolor{black}{encrypted image}. \textcolor{black}{After being embedded with several parameters and the $\overline{\textbf{\textit{SI}}}$ share, the $n$ encrypted images are separately sent to $n$ data hiders.}

\subsubsection{$(r,n)$-Threshold Sharing}
Suppose the size of the image $I$ is $M\times N$. For the $(8-l)$ LSBs of the image, we concatenate two or more pixels to obtain larger elements {\color{black}according to Table~\ref{table.ImaCon}}. Suppose that the concatenated image $I^c$ has the size $M'\times N'$. A permutation is performed to it. To ensure a high security level, the permutation sequence is generated by the encryption key and the sum of the pixel values in image~$I^c$. Then the permuted image can be divided into $\left \lceil \frac{M'\times N'}{r-1} \right \rceil$ sections and each section has $r-1$ pixels. If the last section has less than $r-1$ elements, random integers are generated for padding. Using the $r-1$ pixels $\{x_{0},\cdots,x_{r-2}\}$ in the $j$-th section and one previous sharing result, one can construct $n$ polynomials of $r-1$ degree for the section $j$ as

\begin{equation}
\begin{cases}
\label{eqa.poly1}
f_{j}(q_j)=(\sum_{i=0}^{r-2} x_{i}q_j^i + f_{j-1}(q_{j-1}+p)q_j^{r-1} )\ \mbox{mod}\ F; \\
f_{j}(q_j+1)=(\sum_{i=0}^{r-2} x_{i}(q_j+1)^i  \\ \quad\quad\quad\quad\quad\quad + f_{j-1}(q_{j-1}+p)(q_j+1)^{r-1} ) \ \mbox{mod}\ F; \\
\quad\quad\quad \vdots \\
f_{j}(q_j+n-1)=(\sum_{i=0}^{r-2} x_{i}(q_j+n-1)^i \\ \quad\quad\quad\quad + f_{j-1}(q_{j-1}+p)(q_j+n-1)^{r-1} )\ \mbox{mod}\ F;
\end{cases}
\end{equation}
where $q_j$ is the $j$-th element of $\bf{Q}$, which is a pseudorandom integer sequence generated by the encryption key and $q_j \leq F-n$, and $f_{j-1}(q_{j-1}+p)$ is a randomly selected previous sharing result. The $f_{j-1}(q_{j-1}+p)$ {\color{black}can be any} random integer within range $\left [ 0, F \right )$ when $j=1$.

The $n$ outputs $f_{j}(q_j), f_{j}(q_j+1), ... , f_{j}(q_j+n-1)$ are the $j$-th pixel of the $n$ \textcolor{black}{encrypted images}. Note that each output should be \textcolor{black}{separated into multiple $(8-l)$ bits of pixels that is just opposite with the concatenation strategies in Table.~\ref{table.ImaCon}}. After all the sections are processed, the image constructed by the $(8-l)$ LSBs of the original image are encrypted into $n$ \textcolor{black}{encrypted images}. Finally, fill the $l$ MSBs of the \textcolor{black}{encrypted images} using some random bits to obtain $n$ completely \textcolor{black}{encrypted images}.

\subsubsection{Side Information Embedding}
\label{Section.SIE}
As discussed in Section~\ref{sec.si}, the final side information $\overline{\textbf{\textit{SI}}}$ should be embedded into the $n$ \textcolor{black}{encrypted images before being sent} to the data hider. To equally separate the side information into the $n$ \textcolor{black}{encrypted images}, we also encrypt it into $n$ shares using the CFSS. Only owning $r$ shares, one can recover the final side information.

\textcolor{black}{The final side information $\overline{\textbf{\textit{SI}}}$ consists of the side information \textbf{\textit{SI}} and the bit sequence \textbf{\textit{Tsi}} for pre-processing \textbf{\textit{SI}}. First, divide the side information \textit{\textbf{SI}} into 7-bit streams and transform each 7-bit stream, $b_0b_1b_2b_3b_4b_5b_6$, to decimal value by}
\textcolor{black}{
\begin{equation}
\label{eqa.dec}
d_k =\sum_{i=0}^{6}b_i\cdot 2^i.
\end{equation}}

\textcolor{black}{Thus, the side information \textit{\textbf{SI}} contains $\left \lceil \frac{SI}{7} \right \rceil$ decimal values. The $F$ in sharing the side information is set as 127. Then the elements in \textit{\textbf{SI}} with value 127 should be pre-processed. Thus, the \textit{\textbf{Tsi}} is used to record the location of elements with value 127. By scanning the \textit{\textbf{SI}}, the \textit{\textbf{Tsi}} can be generated as
\begin{equation}
\label{eqa.tsi}
\textit{\textbf{Tsi}}=\begin{cases}
\textit{\textbf{Tsi}}\mathbin\Vert 1 & \quad \mbox{for}\ d_k=127;\\
\textit{\textbf{Tsi}}\mathbin\Vert 0 & \quad \mbox{for}\ d_k=126;\\
\textit{\textbf{Tsi}} & \quad \mbox{otherwise}.
\end{cases}
\end{equation}
Since the elements in the \textit{\textbf{SI}} with values 127 and 126 are only a small probability, the \textit{\textbf{Tsi}} is very short. To avoid the appearance of 127 in \textit{\textbf{Tsi}}, we add a `0' after every six bits into the \textit{\textbf{Tsi}}. Then the maximum value in \textit{\textbf{Tsi}} is 126. Finally, append the \textit{\textbf{Tsi}} to the end of the \textit{\textbf{SI}} to obtain the final $\overline{\textbf{\textit{SI}}}$.}

\textcolor{black}{Now, the total number of decimal values is $\left \lceil \frac{\overline{\textbf{\textit{SI}}}}{7} \right \rceil$ and all the values are smaller than 127. Then divide these decimal values into $\left \lceil \frac{  \overline{\textbf{\textit{SI}}}  }{7(r-1)} \right \rceil$ sections and each section has $r-1$ elements.}
Another pseudorandom integer sequence $\bf{\tilde{Q}}$ of length \textcolor{black}{ $\left \lceil \frac{\overline{\textbf{\textit{SI}}}}{7(r-1)} \right \rceil$ }is generated using the encryption key. All the elements in $\bf{\tilde{Q}}$ aren't larger than \textcolor{black}{ $127-n$ as the modular coefficient is set as $F=127$}. Thus, using the $r-1$ values $\{d_0,\dots,d_{r-2}\}$ in the $k$-th section, and one randomly selected previous sharing result, one can construct $n$ polynomials of $r-1$ degree for the section $k$ as
\textcolor{black}{
\begin{equation}
\begin{cases}
\label{eqa.poly2}
f_{k}(\tilde{q}_k)=(\sum_{i=0}^{r-2} d_{i}\tilde{q}_k^i + f_{k-1}(\tilde{q}_{k-1}+p)\tilde{q}_k^{r-1})\ \mbox{mod} \ 127; \\
f_{k}(\tilde{q}_k+1)=(\sum_{i=0}^{r-2} d_{i}(\tilde{q}_k+1)^i \\ \quad\quad\quad\quad\quad\quad + f_{k-1}(\tilde{q}_{k-1}+p)(\tilde{q}_k+1)^{r-1})\ \mbox{mod} \ 127; \\
\quad\quad\quad \vdots \\
f_{k}(\tilde{q}_k+n-1)=(\sum_{i=0}^{r-2} d_{i}(\tilde{q}_k+n-1)^i \\ \quad\quad\quad\quad + f_{k-1}(\tilde{q}_{k-1}+p)(\tilde{q}_k+n-1)^{r-1})\ \mbox{mod} \ 127,
\end{cases}
\end{equation}}
where $\tilde{q}_k$ is the $k$-th element of $\bf{\tilde{Q}}$, and $f_{k-1}(\tilde{q}_{k-1}+p)$ is a randomly selected previous sharing result. The $f_{k-1}(\tilde{q}_{k-1}+p)$ is a random integer within range $\left [ 0, 127 \right )$ when $k=1$. The $n$ outputs $\{f_{k}(\tilde{q}_k), f_{k}(\tilde{q}_k+1), ... , f_{k}(\tilde{q}_k+n-1)\}$ are the $k$-th element of the $n$ shares. After all the sections are processed, the final side information $\overline{\textbf{\textit{SI}}}$ can be encrypted to be $n$ shares.

{\color{black}The $l$ MSBs of the encrypted images are reversed for data embedding}. After sharing the $\overline{\textbf{\textit{SI}}}$, {\color{black}we embed each $\overline{\textbf{\textit{SI}}}$ share and some additional bits into each encrypted image by substituting its $l$ MSBs. Since each $\overline{\textbf{\textit{SI}}}$ share contains $\left \lceil \frac{\overline{\textbf{\textit{SI}}}}{7(r-1)} \right \rceil$ decimal values. Its length is $\left \lceil \frac{\overline{\textbf{\textit{SI}}}}{7(r-1)} \right \rceil \times7$ bits. Considering the theoretical maximum lengths of $\overline{\textbf{\textit{SI}}}$ and smallest $r$, it is sufficient to use $\left \lceil \log_{2}MN \right \rceil +4$ bits to present the length of each $\overline{\textbf{\textit{SI}}}$ share. First, we use 3 bits, 8 bits, 8 bits and 40 bits to embed the optimal level $l$, parameter $r$, identity of the encrypted image, and the original image size height $M$ and width $N$, respectively, and then use $\left \lceil \log_{2}MN \right \rceil +4$ bits to embed the length of the $\overline{\textbf{\textit{SI}}}$ share. Finally, we embed the $\overline{\textbf{\textit{SI}}}$ share. All these additional bits and $\overline{\textbf{\textit{SI}}}$ share form the overhead \textbf{\textit{OH}} of the encrypted image.}

\subsection{Data Hiding}
\label{sec.demb}
For one data hider who has the $i$-th \textcolor{black}{encrypted image} $E_i$, he/she can directly embed secret data to the encrypted domain without knowing the encryption key or the contents of the original image. The secret data are first encrypted by an existing cryptographic algorithm (e.g. DES or AES~\cite{akkar2001desaes}) using a data hiding key $K_d$. {\color{black}The optimal level $l$ can be extracted from the first three embedded bits. Then, the available embedding space can be found by checking the length of overhead \textbf{\textit{OH}}}. The data hider can embed the encrypted secret data by substituting the $l$ MSBs of the pixels. For example, when $l=3$, the embedding process can be described as
\begin{equation}
\label{eqa.emb}
\textcolor{black}{y_{i,j}' = (m_1\times 2^2 + m_2\times 2^1 + m_3)\times 2^5 + (y_{i,j}\ \mbox{mod}\ 2^5),}
\end{equation}
where $m_1$, $m_2$ and $m_3$ are three bits of the encrypted secret data, $y_{i,j}$ is a pixel of the \textcolor{black}{encrypted image} and the $y_{i,j}'$ is a pixel of the marked encrypted image.

\subsection{Data Extraction and Image Reconstruction}
\textcolor{black}{At least $r$ marked encrypted images are required to completely reconstruct the original image. Since an encryption key is used in the encryption process and a data hiding key is used in the data hiding process, the receiver can extract the embedded data using the data hiding key and reconstruct the original image using the encryption key.}

\subsubsection{Data Extraction with Data Hiding Key}
The CFSS-RDHEI generates $n$ different \textcolor{black}{encrypted images} and the data hider can embed secret data into each \textcolor{black}{encrypted image to generate marked encrypted image}. For each \textcolor{black}{marked encrypted image}, if the receiver has the related data hiding key, he/she can extract the embedded secret data. According to the embedding strategy, all the data are embedded in the $l$ MSBs of the marked encrypted image. Therefore, these embedded data can be extracted from one pixel as follows:
\begin{equation}
\textcolor{black}{\tilde{d}_k= \left \lfloor (y_{i,j}'\ \mbox{mod}\ 2^{9-k}) / 2^{8-k} \right \rfloor,}
\end{equation}
where $k\in\{1,2,\cdots,l\}$ is the $k$-th embedded bit in each pixel.

{\color{black}First, the receiver can obtain the optimal level $l$, parameter $r$, identity of the encrypted image, and the length of the $\overline{\textbf{\textit{SI}}}$ share from the extracted bits. When knowing the length, the receiver can exactly obtain the embedded $\overline{\textbf{\textit{SI}}}$ share. Then the left embedded data are the encrypted secret data.} After all the encrypted secret data are extracted, the secret data can be obtained by decrypting the extracted data using the correct data hiding key.

\subsubsection{Image Reconstruction with Encryption Key}
When \textcolor{black}{owning $r$ marked encrypted images and the encryption key, a receiver} can recover the original image by the following steps.

\begin{itemize}
\item\textit{\textbf{Step 1}}: \textcolor{black}{First, the receiver extracts the optimal level $l$, parameter $r$, identity of each encrypted image, original image height $M$ and width $N$, and $r$ shares of $\overline{\textbf{\textit{SI}}}$ from these marked encrypted images.} Generate the same pseudorandom integer sequence $\bf{\tilde{Q}}$ using the encryption key and then perform the reverse CFSS to recover the final side information $\overline{\textbf{\textit{SI}}}$. {\color{black}Then the side information \textbf{\textit{SI}} can be obtained. Finally,} the compressed location map \textbf{\textit{LM}} and its compression type, $l$-MSB prediction error codes \textbf{\textit{Lpes}}, binary sequence \textbf{\textit{B}}, and reference information \textbf{\textit{T}} can be recovered.

\item \textit{\textbf{Step 2}}: Generate the same pseudorandom integer sequence $\bf{Q}$ using the encryption key. \textcolor{black}{Combine the $(8-l)$ LSBs of the encrypted images according to Table~\ref{table.ImaCon} and perform the reverse CFSS to recover the modified concatenated image $\tilde{I^c}$.}

\item \textit{\textbf{Step 3}}: {\color{black}Recover the original concatenated image $I^c$ using sequence \textbf{\textit{T}}. If a pixel value in $\tilde{I^c}$ is $F-1$, three bits are extracted from \textbf{\textit{T}} when $F=251$ (two bits are extracted when $F=61$) and added to the pixel. After all pixels have been processed, a reverse permutation is followed to obtain the original concatenated image $I^c$. Then, after pixel separating according to Table~\ref{table.ImaCon}}, the $(8-l)$ LSBs of the original image $I$ are reconstructed.

\item \textit{\textbf{Step 4}}: Finally, the MSB prediction is performed to recover the $l$ MSBs of the original image $I$ using the \textbf{\textit{LM}}, \textbf{\textit{Lpes}}, and \textbf{\textit{B}}. The uncompressed location map is obtained from \textbf{\textit{LM}}. The pixels are scanned from left to right and from top to bottom. {\color{black}First, $l$ bits are extracted from \textbf{\textit{B}} to recover the first pixel to initialize the prediction.} Then, for each pixel, if its marker in the location map is `1', {\color{black}it is an unpredicable pixel and its $l$ MSBs are directly extracted from \textbf{\textit{B}}. If its marker in the location map is `0', it is a predicable pixel. We first calculate its prediction value $\tilde{x}_{i,j}^{lMSB}$ using MED predictor as mentioned in Section~\ref{sec.si} and get a prediction error code from the \textbf{\textit{Lpes}}. Then using the $\tilde{x}_{i,j}^{lMSB}$ and error code $\textbf{\textit{Lpes}}_k$, the $l$ MSBs of the original pixel can be recovered as
    \begin{equation}
        \label{eqa.lr}
        x_{i,j}^{lMSB}=\begin{cases}
        \tilde{x}_{i,j}^{lMSB} & \quad \mbox{for}\ \textbf{\textit{Lpes}}_k = 0; \\
        \tilde{x}_{i,j}^{lMSB} - 1 & \quad \mbox{for}\ \textbf{\textit{Lpes}}_k = 10;\\
        \tilde{x}_{i,j}^{lMSB} + 1 & \quad \mbox{for}\ \textbf{\textit{Lpes}}_k = 11.
        \end{cases}
        \end{equation}}
     The original pixel can be reconstructed by combining the recovered $l$ MSBs and $(8-l)$ LSBs. After reconstructing all the pixels, the original image $I$ can be obtained.
\end{itemize}
	
\begin{figure*}[!htbp]
	\centering
	\begin{minipage}[b]{0.87\linewidth}
		\centerline{\includegraphics[width=1\linewidth]{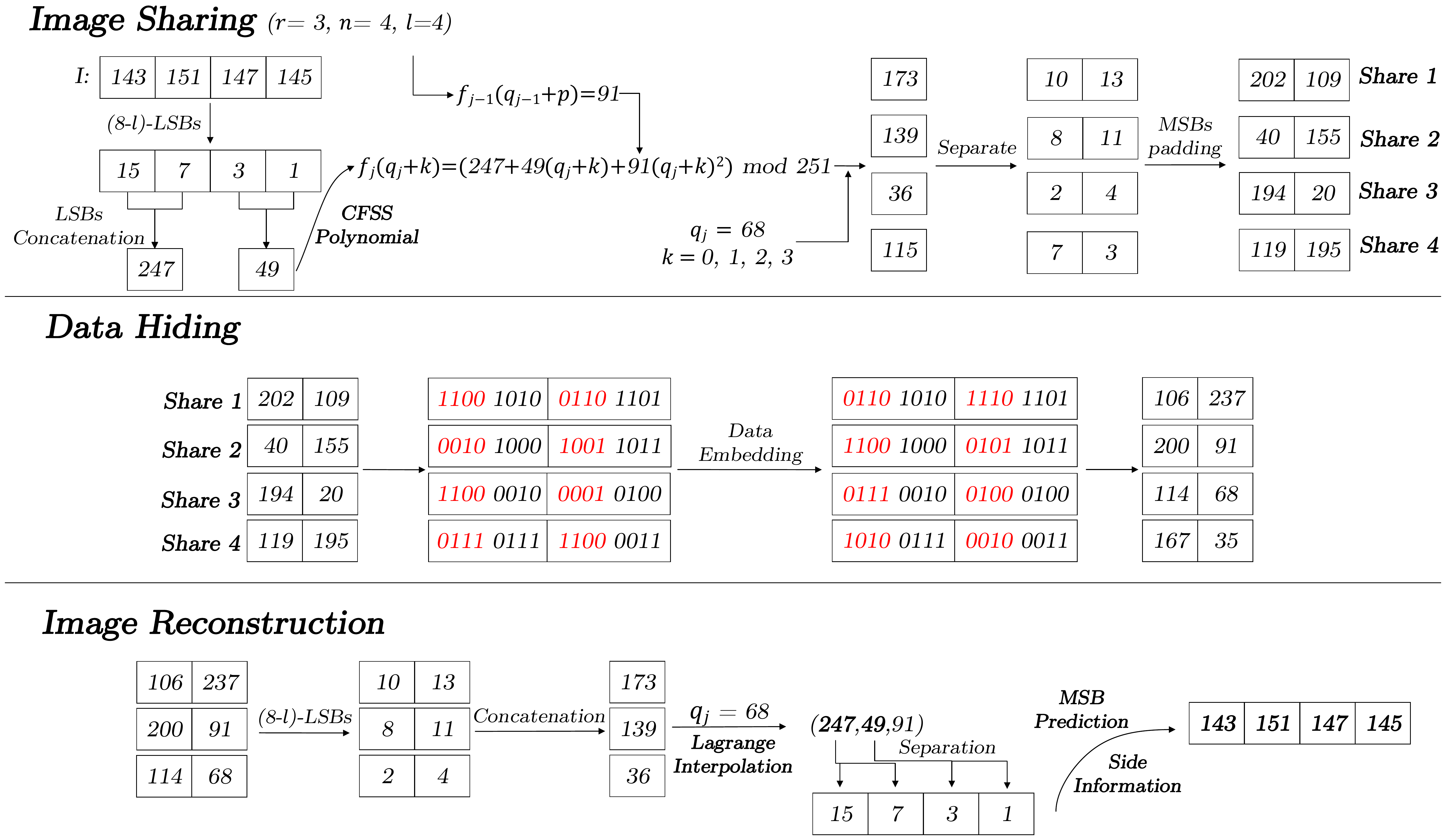}}
	\end{minipage}
	\caption{A numeral example of the CFSS-RDHEI scheme with $(3,4)$-threshold.}
	\label{fig.exp1}
\end{figure*}

\subsection{A Numeral Example}

To better show the main procedures of the proposed CFSS-RDHEI scheme, we provide a numeral example with $(3,4)$-threshold, which indicates that an image is encrypted into \textcolor{black}{four encrypted images} and one can reconstruct the original image using \textcolor{black}{three of them}. Fig.~\ref{fig.exp1} demonstrates the whole operations of the image sharing, data hiding and image reconstruction. Suppose that the optimal level $l$ is 4, {\color{black}then two pixels in the $(8-l)$ LSBs of the image are concatenated according to Table~\ref{table.ImaCon}}. Thus, four pixels of the original image are required in one sharing operation and they are supposed as $(143,151,147,145)$ in the $j$-th section. Besides, the $j$-th element in the pseudorandom integer sequence $\bf{Q}$ is $q_j= 68$.

First, we get the $(8-l)$-LSB values of these four pixels {\color{black}as $(15,7,3,1)$} and then concatenate them to two elements $\hat{x}_0 = $ 247 and $\hat{x}_1 =$ 49. Using $\hat{x}_0 $, $\hat{x}_1$, $q_j$, and one previous sharing result {\color{black}$f_{j-1}(q_{j-1} +p)=91$}, one can get the secret sharing polynomials
{\color{black}
\begin{equation}
\label{eqa.poly}
f_{j}(q_j+k)=(\hat{x}_{0}+\hat{x}_{1}(q_j+k)+f_{j-1}(q_{j-1} +p)(q_j+k)^2)\ \mbox{mod}\ F, \\
\end{equation}
where $k\in\{0,1,2,3\}$.
}

For the modular coefficient $F=251$, the four outputs $f_j(q_j), f_j(q_j+1), f_j(q_j+2)$ and $f_j(q_j+3)$ can be obtained and separated to be the $(8-l)$-LSBs of two pixels in each \textcolor{black}{encrypted image}, which are $(10,13),(8,11),(2,4)$ and $(7,3)$, respectively. The $l$ MSBs of these pixels are filled using random bits. Then four \textcolor{black}{encrypted images} can be obtained and they are $(202,109),(40,155),(194,20)$, and $(119,195)$. Each data hider can hide $l$ bits of secret data into the $l$ MSBs of each pixel. As show from Fig.~\ref{fig.exp1}, the four images become $(106,237),(200,91),(114,68)$, and $(167,35)$ after embedding secret data into their $l$ MSBs.

Suppose that the three \textcolor{black}{marked encrypted images} with pixels $(106,237)$, $(200,91)$, and $(114,68)$ are collected. First, calculate the $(8-l)$-LSBs of the three shares as $(10,13),(8,11)$ and $(2,4)$. Then they are concatenated to three elements as $(173,139,36)$. Perform the reverse CFSS using the same integer $q_j=68$ to recover the $(8-l)$-LSBs of the four pixels as $(15,7,3,1)$. Finally, perform the MED predictor with the side information, the original pixels can be recovered.

\subsection{Discussion}
Compared with other secret sharing-based RDH-EI methods, the proposed CFSS-RDHEI has the following advantages.
\begin{enumerate}
\item The CFSS-RDHEI can encrypt an original image to be $n$ \textcolor{black}{encrypted images} following the cryptography standards such that the encrypted images have high security.
\item The CFSS-RDHEI has much lower data expansion than other similar methods and each \textcolor{black}{encrypted image has only} the $1/(r-1)$ size of the original image.
\item Since {\color{black}any random integer} can be used in sharing the first section, the CFSS-RDHEI is a non-deterministic system. When encrypting an image using a same encryption key and parameters $n$ and $r$, the generated \textcolor{black}{encrypted images} may be completely different and unpredictable, which provides a strong ability to resist {\color{black}many potential} attacks.
\item A multi-MSBs prediction method {\color{black}with secure location map} is used to embed secret data and it can overcome the security problems in most existing MSB prediction methods.
\item The CFSS-RDHEI can obtain a large embedding rate for many natural images.

\end{enumerate}

\section{Simulation Results and Performance Analysis}
\label{Section4}
\begin{figure}[!htbp]
	\centering
	\begin{minipage}[b]{0.99\linewidth}
		\begin{minipage}[b]{0.24\linewidth}
			\centerline{\includegraphics[width=1\linewidth]{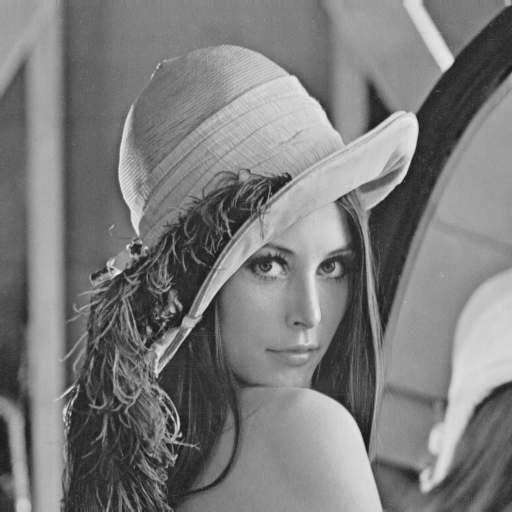}}
			\centerline{(a)}
		\end{minipage}\hfill
		\begin{minipage}[b]{0.24\linewidth}
			\centerline{\includegraphics[width=1\linewidth]{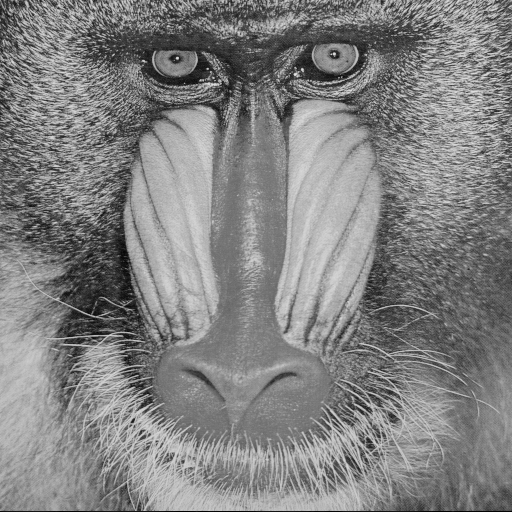}}
			\centerline{(b)}
		\end{minipage}\hfill
		\begin{minipage}[b]{0.24\linewidth}
			\centerline{\includegraphics[width=1\linewidth]{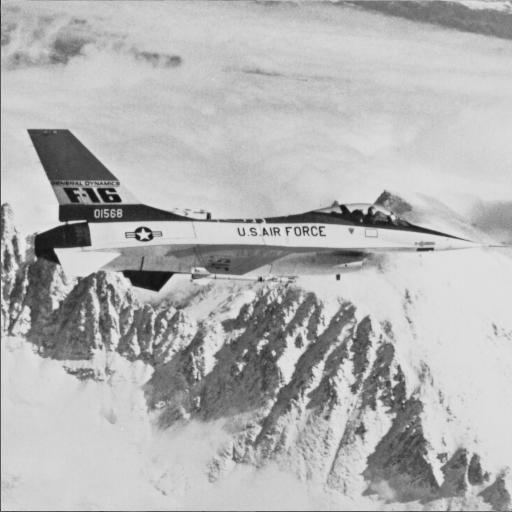}}
			\centerline{(c)}
		\end{minipage}\hfill
		\begin{minipage}[b]{0.24\linewidth}
			\centerline{\includegraphics[width=1\linewidth]{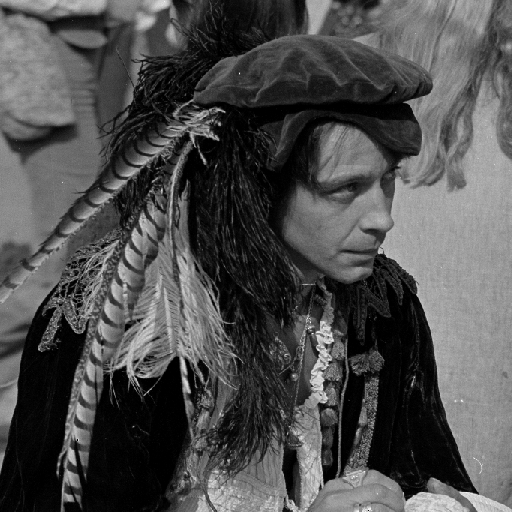}}
			\centerline{(d)}
		\end{minipage}\hfill\\
		\begin{minipage}[b]{0.24\linewidth}
			\centerline{\includegraphics[width=1\linewidth]{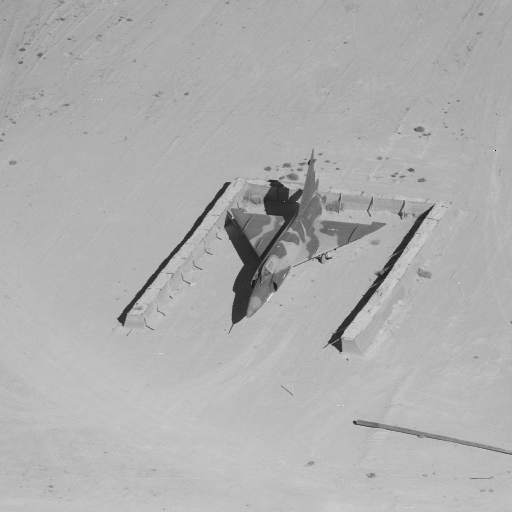}}
			\centerline{(e)}
		\end{minipage}\hfill
		\begin{minipage}[b]{0.24\linewidth}
			\centerline{\includegraphics[width=1\linewidth]{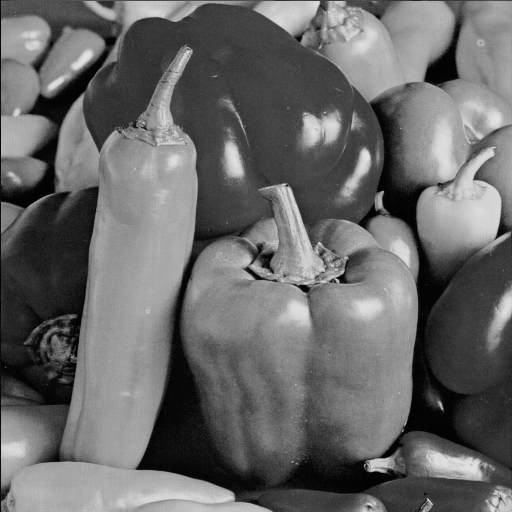}}
			\centerline{(f)}
		\end{minipage}\hfill
		\begin{minipage}[b]{0.24\linewidth}
			\centerline{\includegraphics[width=1\linewidth]{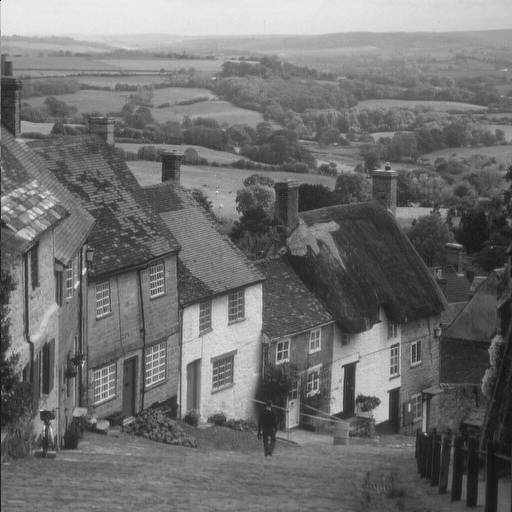}}
			\centerline{(g)}
		\end{minipage}\hfill
		\begin{minipage}[b]{0.24\linewidth}
			\centerline{\includegraphics[width=1\linewidth]{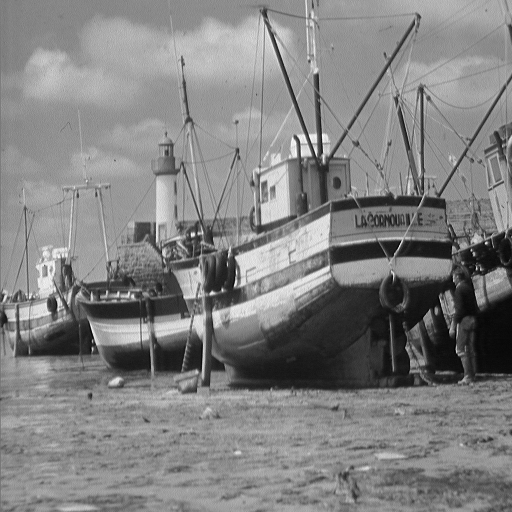}}
			\centerline{(h)}
		\end{minipage}\hfill
	\end{minipage}\hfill
	\caption{The eight test images of size $512\times 512$. (a) \textit{Lena}; (b) \textit{Baboon}; (c) \textit{Jetplane}; (d) \textit{Man}; (e) \textit{Airplane}; (f) \textit{Peppers}; (g) \textit{Goldhill}; (h) \textit{Boat}.}
	\label{fig.testimg}
\end{figure}

In this section, we simulate the CFSS-RHDEI scheme, evaluate its performance, and compare it with {\color{black}existing secret sharing-based RDH-EI} schemes. Eight commonly used greyscale images of size $512\times 512$ are selected as the test images and they are shown in Fig.~\ref{fig.testimg}. The secret data to be embedded are randomly generated. \textcolor{black}{They are encrypted using some encryption standards such as AES before being embedded. These standards can encrypt the secret data into encrypted data with high randomness and security}.

\subsection{Simulation Results}
\label{Section.SimuRes}

\begin{figure}[!htbp]
	\centering
	\begin{minipage}[b]{0.99\linewidth}
		\begin{minipage}[b]{0.32\linewidth}
			\centerline{\includegraphics[width=1\linewidth]{graylena.png}}
			\centerline{(a)}
		\end{minipage}\hfill
		\begin{minipage}[b]{0.32\linewidth}
			\centerline{\includegraphics[width=1\linewidth]{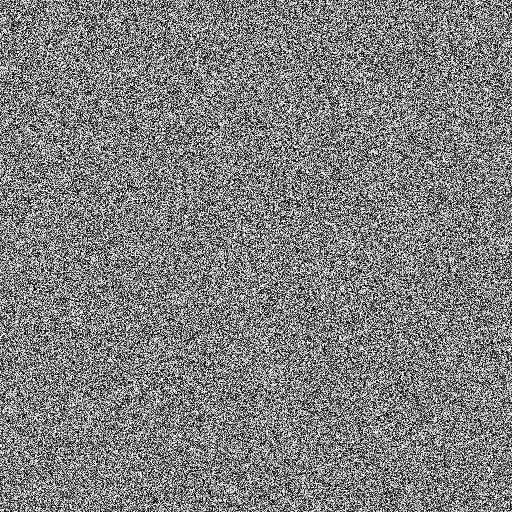}}
			\centerline{(b)}
		\end{minipage}\hfill
		\begin{minipage}[b]{0.32\linewidth}
			\centerline{\includegraphics[width=1\linewidth]{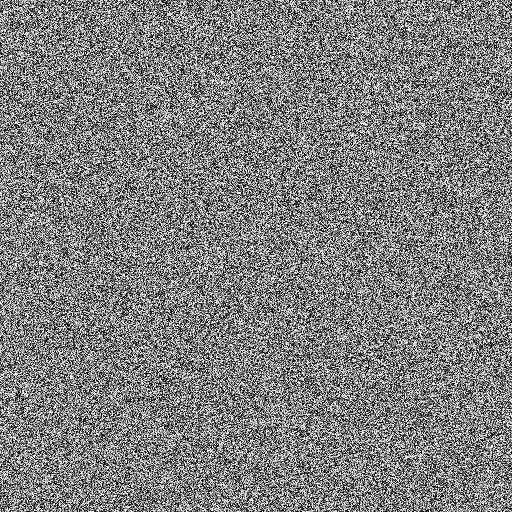}}
			\centerline{(c)}
		\end{minipage}\\
		\begin{minipage}[b]{0.32\linewidth}
			\centerline{\includegraphics[width=1\linewidth]{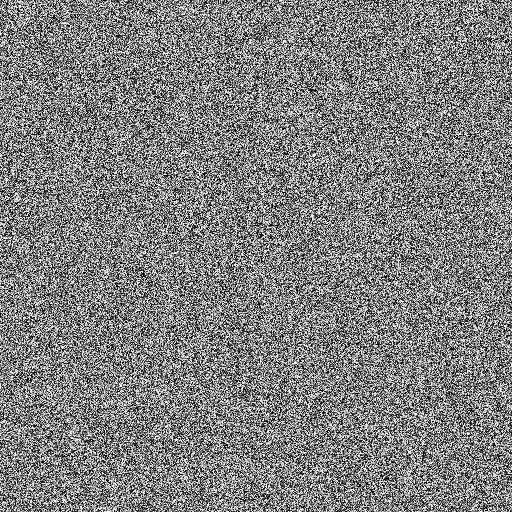}}
			\centerline{(d)}
		\end{minipage}\hfill
		\begin{minipage}[b]{0.32\linewidth}
			\centerline{\includegraphics[width=1\linewidth]{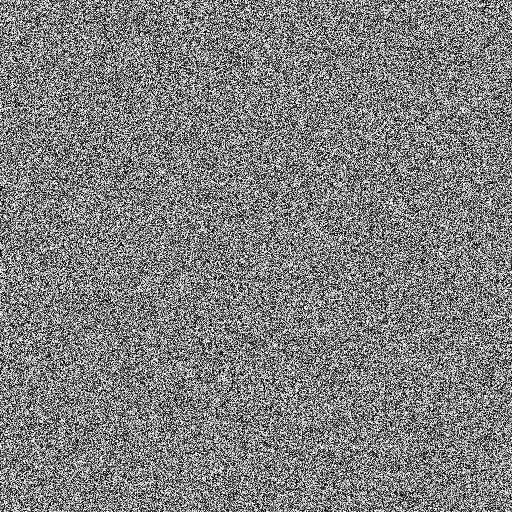}}
			\centerline{(e)}
		\end{minipage}\hfill
		\begin{minipage}[b]{0.32\linewidth}
			\centerline{\includegraphics[width=1\linewidth]{graylena.png}}
			\centerline{(f)}
		\end{minipage}\hfill
	\end{minipage}\hfill
	\caption{Simulation results of the CFSS-RDHEI with $(2,2)$-threshold image sharing. (a) Original Image \textit{Lena} of size~$512\times 512$; (b)-(c) \textcolor{black}{Two encrypted images}; (d)-(e) \textcolor{black}{Two marked images}; (f) Reconstructed Image \textit{Lena}.}
	\label{fig.explena}
\end{figure}

Fig.~\ref{fig.explena} shows the simulation results of the $(2,2)$-threshold image sharing for the test image $Lena$. According to Eq.~\eqref{equ.thres}, the optimal level for the image $Lena$ is $l=5$. \textcolor{black}{Fig.~\ref{fig.explena}(a) is the original $Lena$ image and Figs.~\ref{fig.explena}(b) and (c) are two encrypted images generated by the CFSS. It can be seen that all the encrypted images are noise-like and don't contain any information of the original information. Figs.~\ref{fig.explena}(d) and (e) are two marked encrypted images obtained by embedding secret data into Figs.~\ref{fig.explena}(b) and (c), respectively. Since all the secret data are encrypted into random data before embedding, the marked encrypted images won't reveal any information of the original image or the secret data. Fig.~\ref{fig.explena}(f) is the reconstructed image obtained from Figs.~\ref{fig.explena}(d) and (e) and the encryption key. It is exactly the same with the original image in Fig.~\ref{fig.explena}(a). So, our CFSS-RDHEI is an lossless scheme.}

\begin{figure}[!htbp]
	\centering
	\begin{minipage}[b]{0.98\linewidth}
		\begin{minipage}[b]{0.2\linewidth}
			\centerline{\includegraphics[width=1\linewidth]{graybaboon.png}}
			\centerline{(a)}
		\end{minipage}\hfill
		\begin{minipage}[b]{0.2\linewidth}
			\centerline{\includegraphics[width=0.5\linewidth]{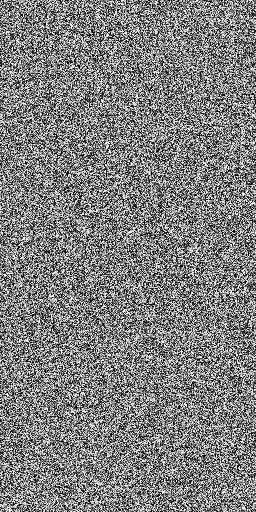}}
			\centerline{(b)}
		\end{minipage}\hfill
		\begin{minipage}[b]{0.2\linewidth}
			\centerline{\includegraphics[width=0.5\linewidth]{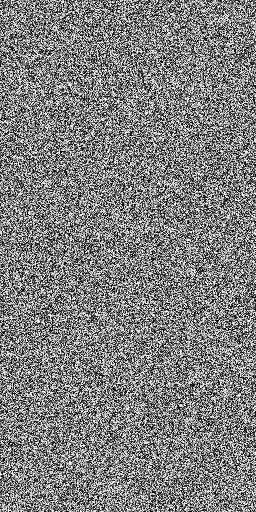}}
			\centerline{(c)}
		\end{minipage}\hfill
		\begin{minipage}[b]{0.2\linewidth}
			\centerline{\includegraphics[width=0.5\linewidth]{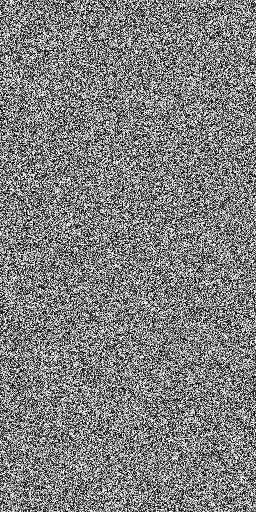}}
			\centerline{(d)}
		\end{minipage}\hfill
		\begin{minipage}[b]{0.2\linewidth}
			\centerline{\includegraphics[width=0.5\linewidth]{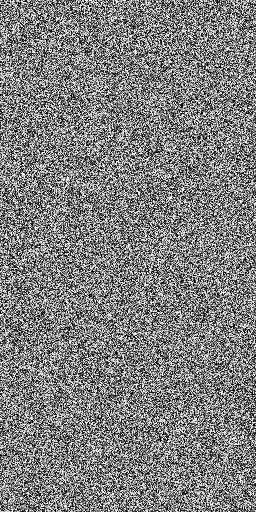}}
			\centerline{(e)}
		\end{minipage}\\
		\begin{minipage}[b]{0.2\linewidth}
			\centerline{\includegraphics[width=0.5\linewidth]{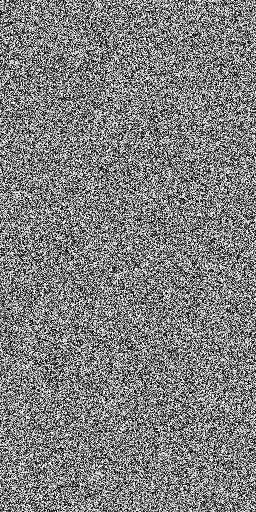}}
			\centerline{(f)}
		\end{minipage}\hfill
		\begin{minipage}[b]{0.2\linewidth}
			\centerline{\includegraphics[width=0.5\linewidth]{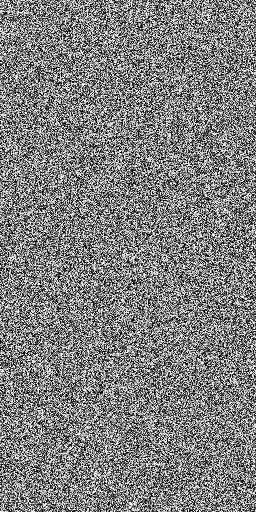}}
			\centerline{(g)}
		\end{minipage}\hfill
		\begin{minipage}[b]{0.2\linewidth}
			\centerline{\includegraphics[width=0.5\linewidth]{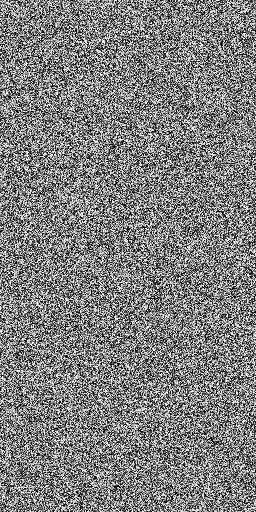}}
			\centerline{(h)}
		\end{minipage}\hfill
		\begin{minipage}[b]{0.2\linewidth}
			\centerline{\includegraphics[width=0.5\linewidth]{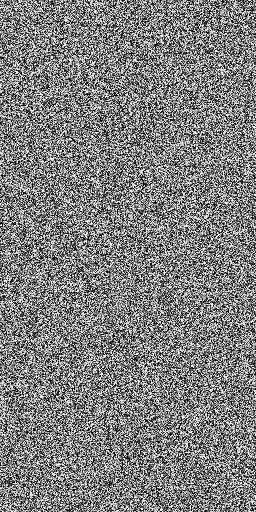}}
			\centerline{(i)}
		\end{minipage}\hfill
		\begin{minipage}[b]{0.2\linewidth}
			\centerline{\includegraphics[width=1\linewidth]{graybaboon.png}}
			\centerline{(j)}
		\end{minipage}\hfill
	\end{minipage}\hfill
	\caption{Simulation results of the CFSS-RDHEI with $(3,4)$-threshold image sharing. (a) Original Image \textit{Baboon} of size~$512\times 512$; (b)-(e) \textcolor{black}{four encrypted images}; (f)-(i) \textcolor{black}{our marked images}; (j) reconstructed image \textit{Baboon} from marked images (f), (g), and (i).}
	\label{fig.expbaboon}
\end{figure}

Fig.~\ref{fig.expbaboon} depicts the simulation results of the $(3,4)$-threshold image sharing for the test image $Baboon$.
\textcolor{black}{Four encrypted images Figs.~\ref{fig.expbaboon}(b), (c), (d), and (e) are generated from the original image Fig.~\ref{fig.expbaboon}(a), and four marked encrypted images Figs.~\ref{fig.expbaboon}(f), (g), (h), and (i) are obtained by embedding data into the four encrypted images. Fig.~\ref{fig.expbaboon}(j) is the reconstructed image from the marked encrypted images Figs.~\ref{fig.expbaboon}(f), (g) and (i). It shows that each marked encrypted image is $1/(r-1)$} of the original image. Thus, when $r>2$, the marked encrypted image has smaller size than the original image.

\subsection{Data Expansion}
{\color{black}Since the encrypted results of homomorphic encryption are larger than the original data and multiple images are generated in a secret share scheme, the total size of the marked encrypted images is often larger than the original image in RDH-EI schemes using homomorphic encryption and secret sharing. According to the discussions in~\cite{wu2018adopting}, the data expansion for the whole scheme is the ratio between the bits of all the marked encrypted images and the bits of the original image, while the data expansion for each data hider is the ratio between the bits of each marked encrypted image and the bits of the original image. The data expansion is expected to be small to save the storage room. The data expansion for each data hider in traditional RDH-EI schemes is 1.}

In our CFSS-RDHEI with $(r,n)$-threshold, a secret image is encrypted into $n$ \textcolor{black}{encrypted images} and each \textcolor{black}{encrypted image} has the $1/(r-1)$ size of the original image. Thus, the total size of the $n$ \textcolor{black}{encrypted images} are the $n/(r-1)$ {\color{black}size} of the original image. Since each data hider keeps one \textcolor{black}{encrypted image}, the expansion rate for each data hider is $1/(r-1)$ size of the original image. Table~\ref{table.expand} lists the data expansion comparison of different secret sharing-based RDH-EI schemes. {\color{black}It can be seen that the expansion rate for the whole scheme is $n$ in the Wu $et~al.$~\cite{wu2018adopting} and Chen $et~al.$~\cite{chen2020secret} methods, and is $n/(r-1)$ in our CFSS-RDHEI. Note that the expansion rate for the whole scheme is always 1 in the Chen $et~al.$~\cite{chen2019newsecret} method, because it only has one encrypted image. Besides, the expansion rate for each data hider is only $1/(r-1)$ in the CFSS-RDHEI, which is much smaller than the other three methods. In general, our proposed method has much less data expansion than other secret sharing-based RDH-EI methods.}

\begin{table}[!htbp]
\small
	\renewcommand{\arraystretch}{1}
	\setlength{\tabcolsep}{8pt}
	\begin{center}
		\caption{Data expansion comparison of the secret sharing-based RDH-EI schemes.}
		\label{table.expand}
		\begin{tabular}{cc c c}
			\hline
			Methods  & \tabincell{c}{Expansion rate for \\ the whole scheme}&  \tabincell{c}{Expansion rate for \\ each data hider} \\
			\hline
			Chen $et~al.$~\cite{chen2019newsecret}&    $1$ &  $1$  \\
			Wu $et~al.$~\cite{wu2018adopting}&     $n$ &  $1$ \\	
			Chen $et~al.$~\cite{chen2020secret}&   $n $ &  $ 1$ \\
			CFSS-RDHEI &    $ n/(r-1) $ &  $1/(r-1)$ \\
			\hline
		\end{tabular}
	\end{center}
\end{table}

\subsection{Embedding Rate}
{\color{black}The effective embedding capacity indicates the maximum number of secret data that can be embedded in the encrypted image by the data hider.} It is usually evaluated by the embedding rate $ER$. {\color{black}The $ER$ is calculated as,}

{\color{black}
\begin{equation}
\label{eqa.rate0}
ER =  \frac{\mbox{Embeddable capacity - Overhead}}{\mbox{Total number of pixels}}.
\end{equation}
The embedding rate is expected to be large to embed more secret data. In our CFSS-RDHEI scheme, the embeddable capacity for an image of size $M\times N$ is $lMN$, where $l$ is the optimal level of the original image, and the overhead \textbf{\textit{OH}} consists of the share of final side information and some additional bits, discussed in Section~\ref{Section.SIE}. Thus, the embedding rate of our CFSS-RDHEI schem is $ER=(lMN- \textbf{\textit{OH}})/{MN}$.}

Table~\ref{table.er} {\color{black}lists the sizes of the embeddable capacity, overhead and effective embedding capacity, and the} embedding rates of our CFSS-RDHEI in eight test images with $(2,n)$-threshold scheme. {\color{black}The optimal level $l$ in different images may be different to result in different embeddable capacity and embedding rates.} For example, the embedding rate of image \textit{Lena} is $2.91\ bpp$ with $l=5$. The embedding rate {\color{black}of an image is} mainly determined by the prediction errors of the image. The images \textit{Baboon} and \textit{Man} have relatively low embedding rates, because they have {\color{black}small $l$ and many unpredictable pixels, which lead to small embeddable capacity and large overhead.} On the contract, the images \textit{Airplane} and \textit{Jetplane} can achieve much larger embedding rates.

\begin{table}[!htbp]
\small
	\renewcommand{\arraystretch}{1}
	\setlength{\tabcolsep}{2.5pt}
	\begin{center}
		\caption{The embedding rates of eight test images with $(2,n)$-threshold.}
		\label{table.er}
		\begin{tabular}{c c >{\color{black}}c >{\color{black}}c>{\color{black}} c c}
			\hline
			\tabincell{c}{Test \\images}  & \tabincell{c}{ Optimal \\ level $l$}  & \tabincell{c}{Embeddable \\ capacity} & \tabincell{c}{ Overhead  \\ \textbf{\textit{OH}}} & \tabincell{c}{Effective embe-\\dding capacity} & \tabincell{c}{\textit{ER} \\ ($bpp$)}   \\\hline
			\textit{Lena} & 5 & 1,310,720  & 547,116 & 763,604 & 2.91 \\
			\textit{Baboon} & 4 & 1,048,576 & 720,307 & 328,269 & 1.25  \\
			\textit{Jetplane} & 6 & 1,572,864 & 723,111 & 849,753 & 3.24  \\
			\textit{Man} & 5 & 1,310,720 & 737,586 & 573,134 & 2.19  \\
			\textit{Airplane} & 6 & 1,572,864 & 619,034 & 953,830 & 3.64  \\
			\textit{Peppers} & 5 & 1,310,720 & 637,113 & 673,607 & 2.57  \\
			\textit{Goldhill} & 5 & 1,310,720 & 660,377 & 650,343 & 2.48  \\
			\textit{Boat} & 5 & 1,310,720 & 582,273 & 728,447 & 2.78  \\
			\hline
		\end{tabular}
	\end{center}
\end{table}

\begin{figure*}[!htbp]
	\centering
	\begin{minipage}[b]{0.9\linewidth}
		\begin{minipage}[b]{0.32\linewidth}
			\centerline{\includegraphics[width=1\linewidth]{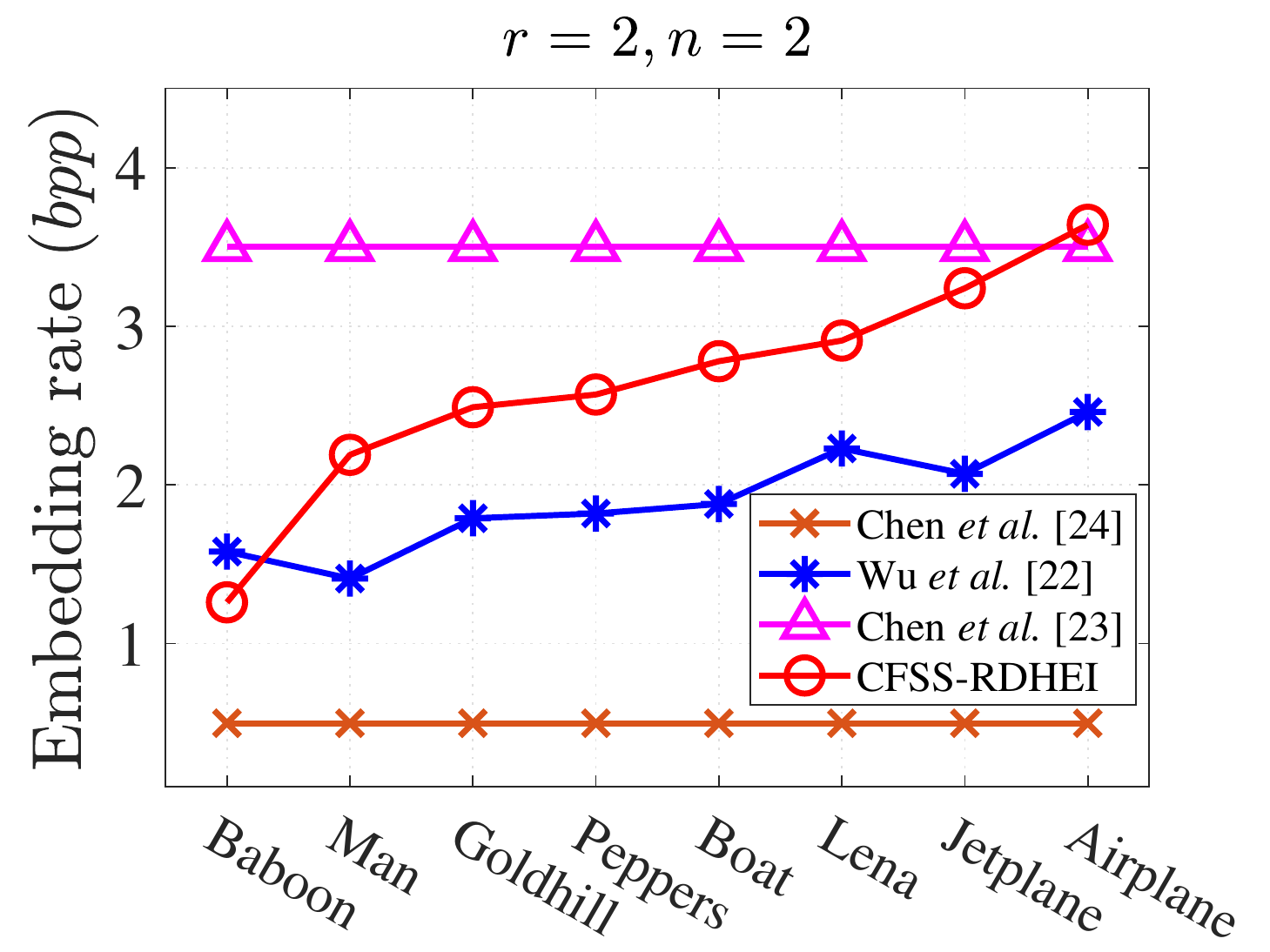}}
			\centerline{(a)}
		\end{minipage}\hfill
		\begin{minipage}[b]{0.32\linewidth}
			\centerline{\includegraphics[width=1\linewidth]{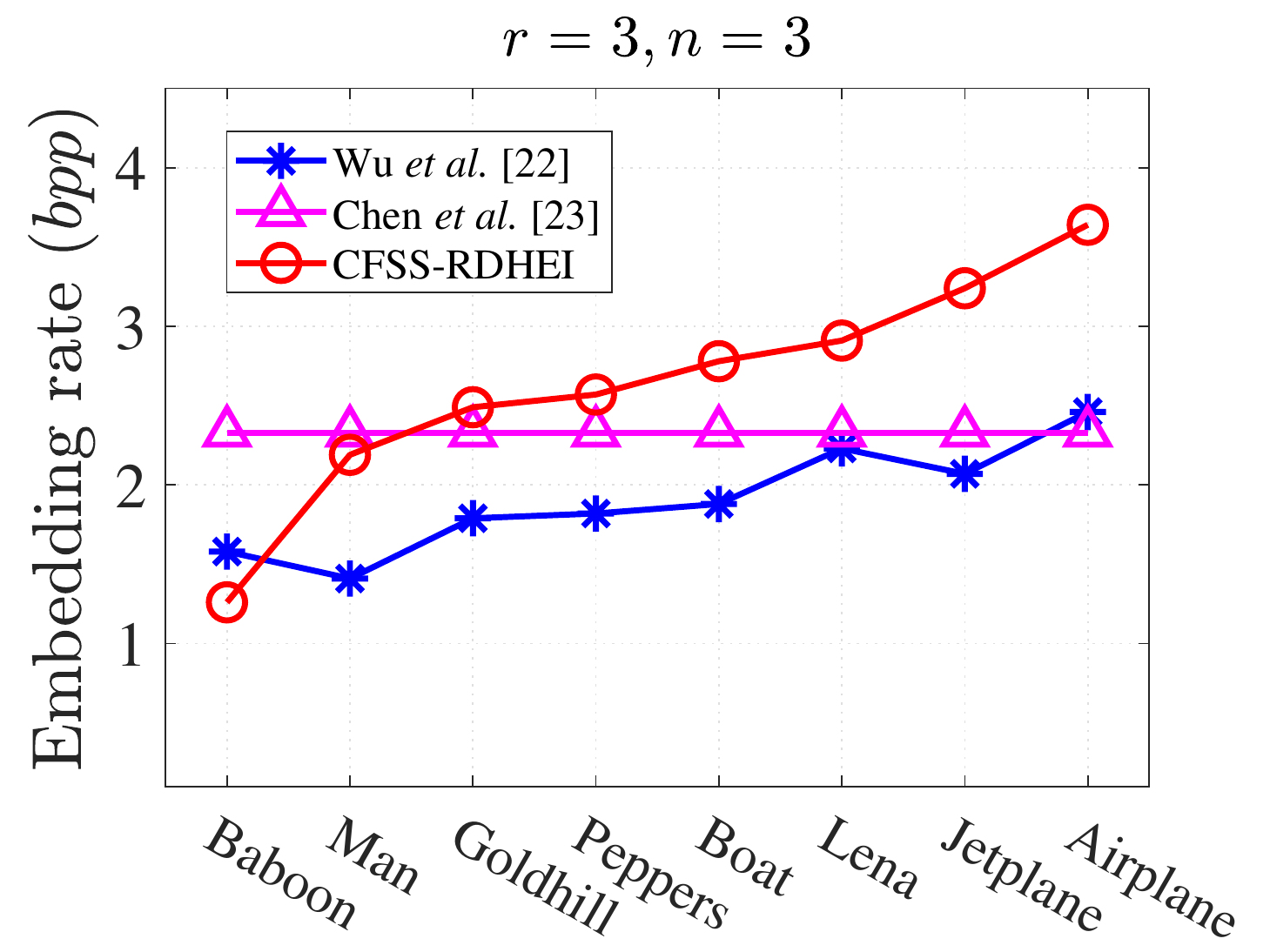}}
			\centerline{(b)}
		\end{minipage}\hfill
		\begin{minipage}[b]{0.32\linewidth}
			\centerline{\includegraphics[width=1\linewidth]{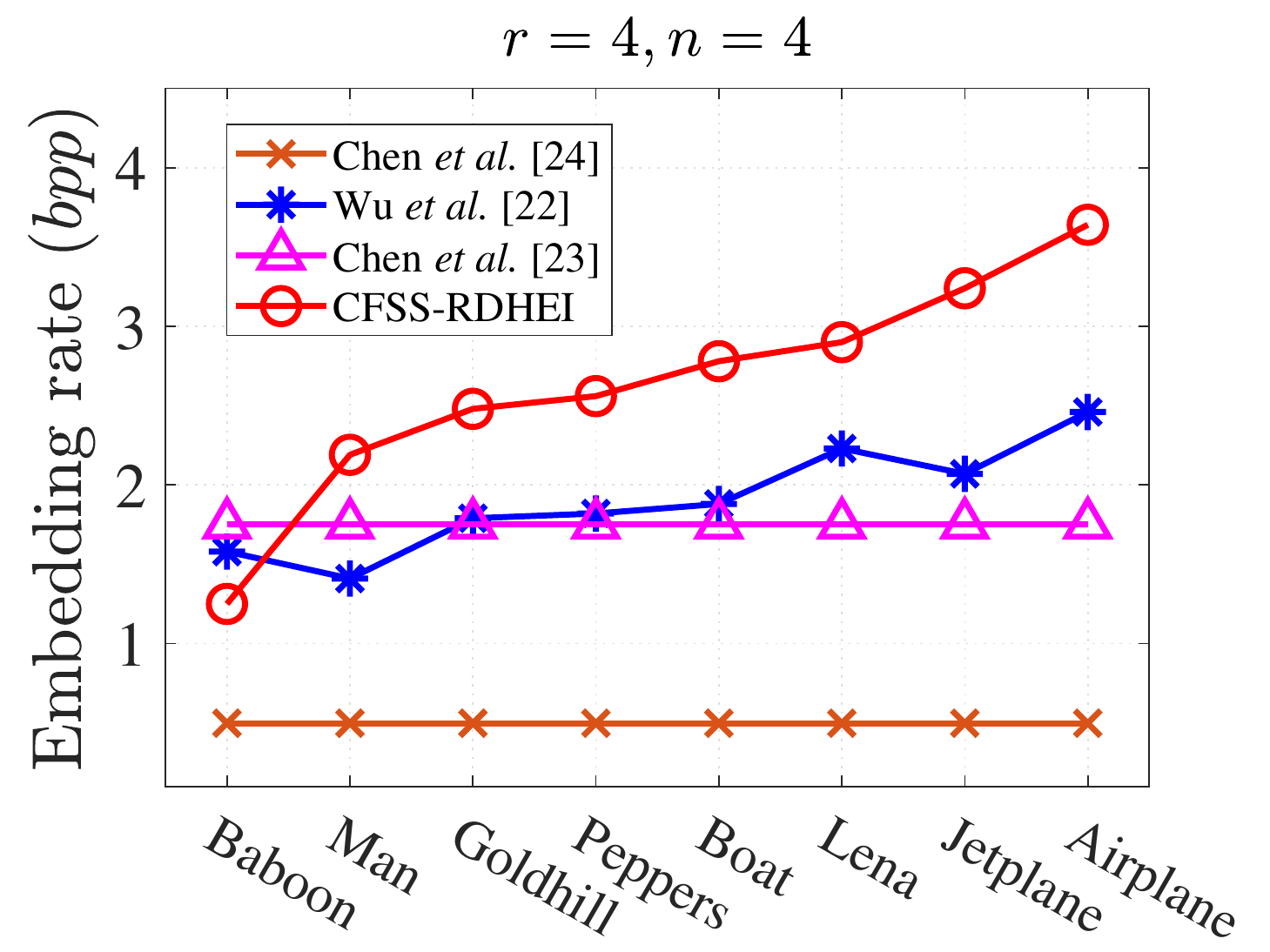}}
			\centerline{(c)}
		\end{minipage}\hfill\\
		\begin{minipage}[b]{0.32\linewidth}
			\centerline{\includegraphics[width=1\linewidth]{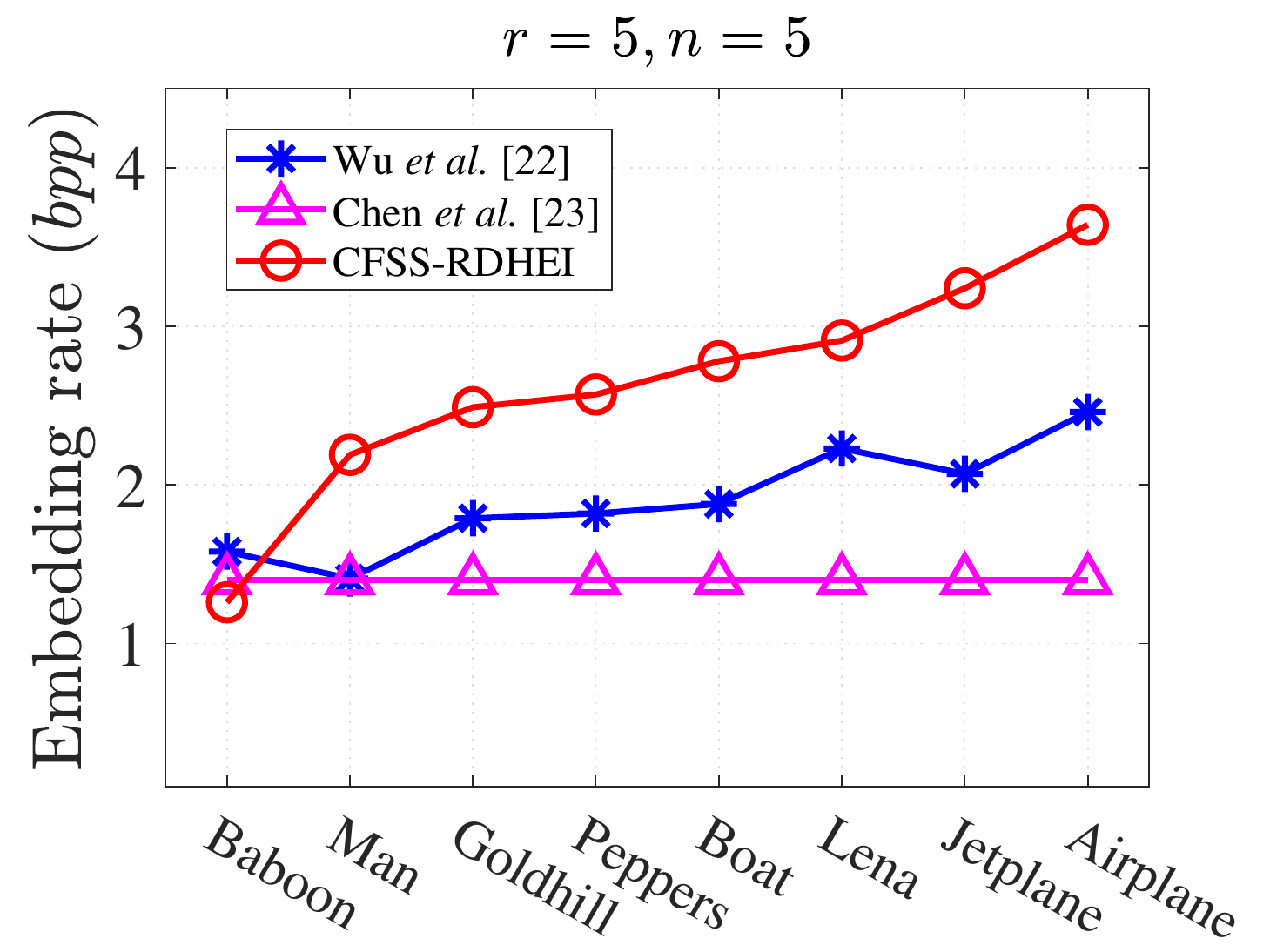}}
			\centerline{(d)}
		\end{minipage}\hfill
		\begin{minipage}[b]{0.32\linewidth}
			\centerline{\includegraphics[width=1\linewidth]{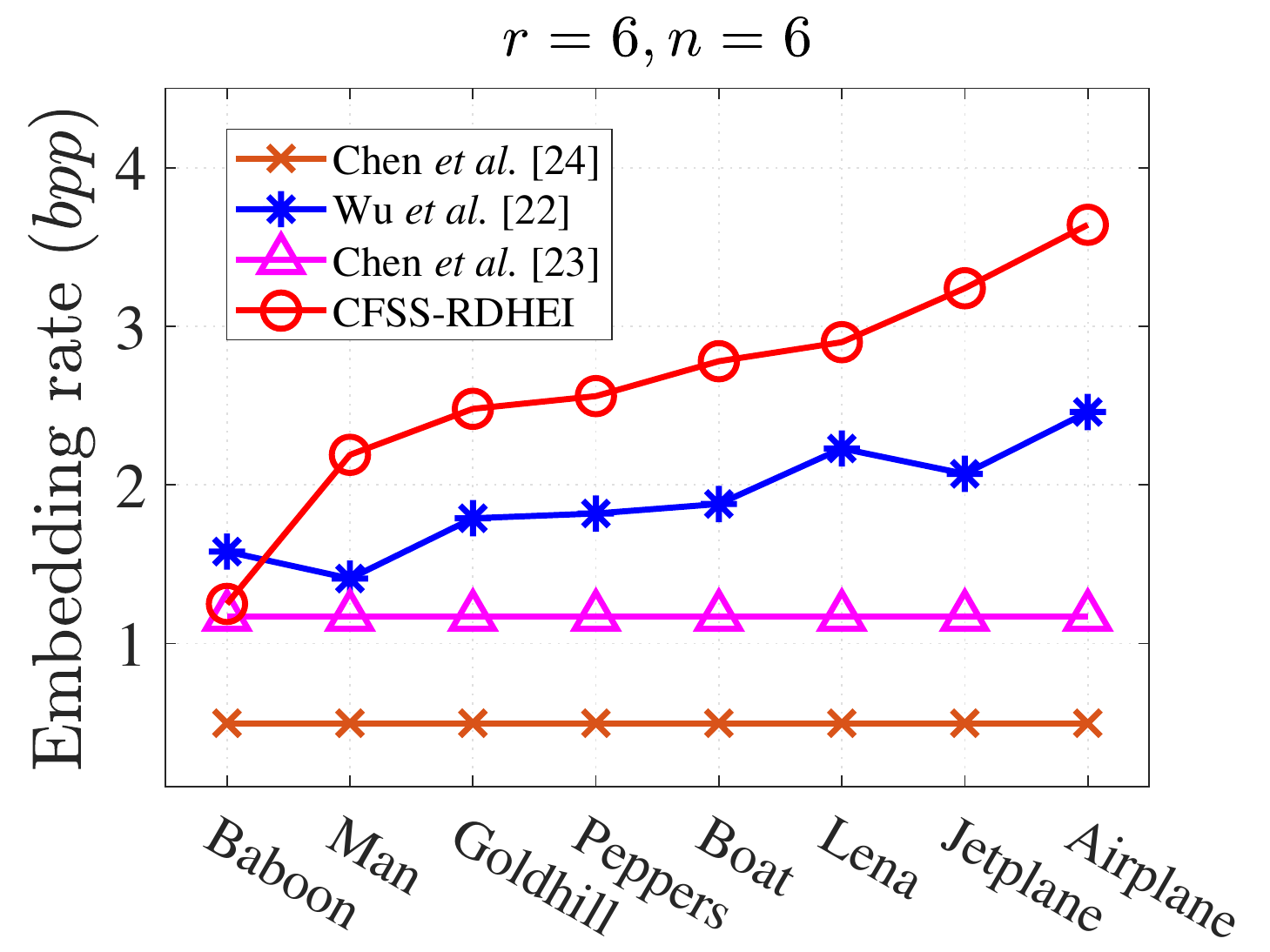}}
			\centerline{(e)}
		\end{minipage}\hfill
		\begin{minipage}[b]{0.32\linewidth}
			\centerline{\includegraphics[width=1\linewidth]{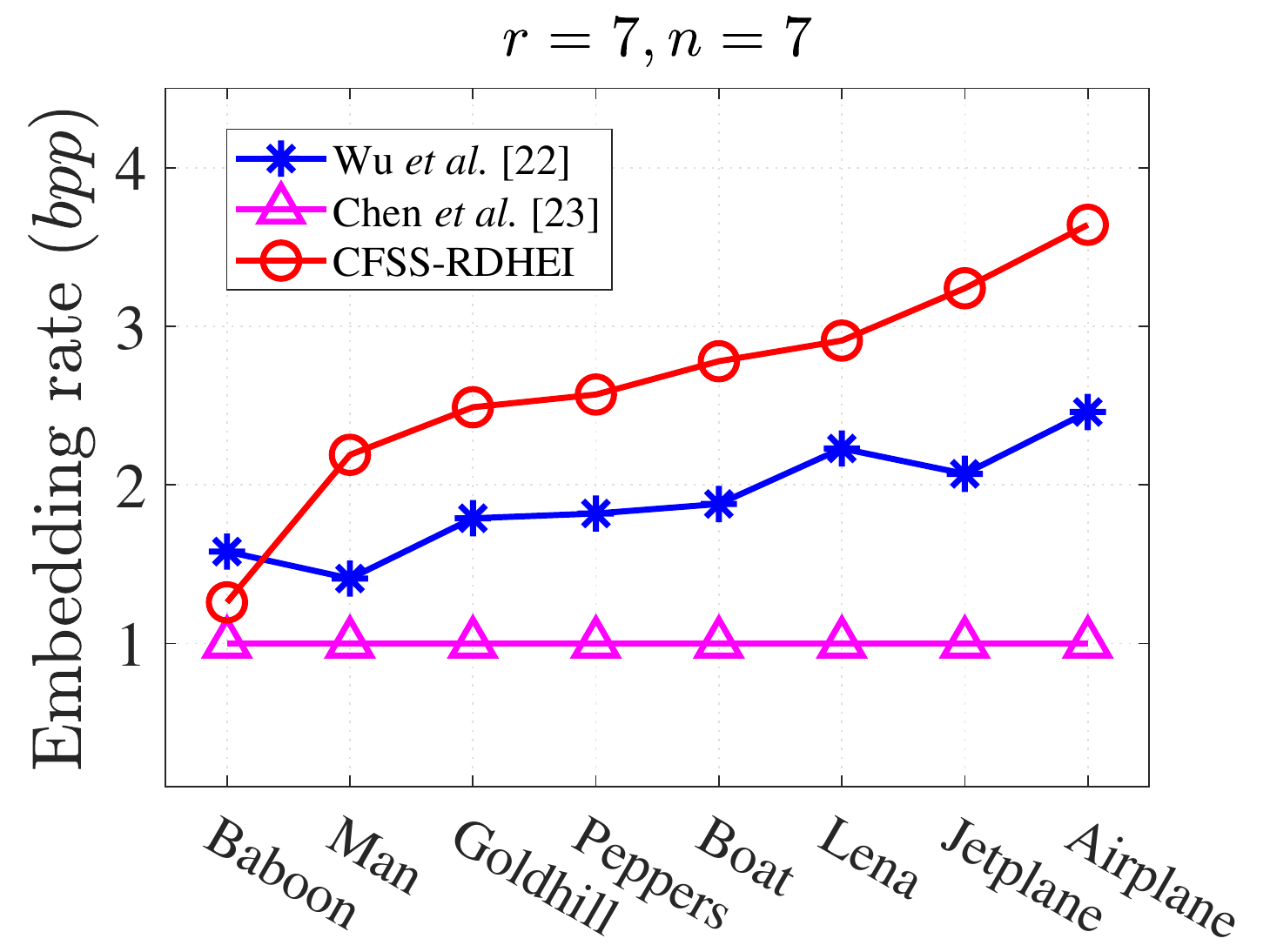}}
			\centerline{(f)}
		\end{minipage}\hfill
	\end{minipage}\hfill
	\caption{Embedding rates of different secret sharing-based RDH-EI schemes at different $(r,n)$-threshold.}
	\label{fig.ebr}
\end{figure*}

For an $(r,n)$-threshold secret share-based RDH-EI, its embedding rate may be determined by the parameters $r$ or $n$. Our experiment compares the embedding rates of the proposed CFSS-RDHEI with the Wu $et~al.$~\cite{wu2018adopting}, Chen $et~al.$~\cite{chen2019newsecret} and Chen $et~al.$~\cite{chen2020secret} methods and Fig.~\ref{fig.ebr} lists the experimental results. As can be seen, the embedding rates of the Wu $et~al.$~\cite{wu2018adopting} and Chen $et~al.$~\cite{chen2019newsecret} methods stay the same with different $r$ or $n$, and the embedding rate of the Chen $et~al.$~\cite{chen2020secret} method is $7/n$. For our CFSS-RDHEI, the embedding rates {\color{black}of an image are slightly different} with different $r$, because {\color{black}the sizes of overhead may be different for different $r$}. The results show that the Chen $et~al.$~\cite{chen2020secret} method can achieve the best embedding rate when $n=2$. It is obvious that with the increase of $r$ or $n$, our CFSS-RDHEI can achieve the best embedding rates for most test images.

\section{Security Analysis}
\label{Section5}

An RDH-EI scheme is expected to have high security to protect the embedded data and the original images. However, most existing RDH-EI schemes can achieve high security only for the embedded data. They cannot well protect the original images. This is because they usually encrypt the original images using some lightweight encryption methods such xor operation, to keep redundant information for data embedding. Thus, the encrypted images cannot defense many security attacks such as the differential and chosen-plaintext attacks. For many RDH-EI schemes using homomorphic encryption, they can achieve high security. However, their data expansion is large and computation cost is high.

In our CFSS-RDHEI, the CFSS strictly follows the cryptography standards and the embedded secret data can be encrypted by the well-known AES~\cite{akkar2001desaes} or any encryption method. Thus, it is able to achieve a high security level for both the embedded secret data and original images. This section evaluates the security of the original image {\color{black}in CFSS-RDHEI} and compares it with {\color{black}existing} secret sharing-based RDH-EI schemes. 

\subsection{Key Sensitivity Analysis}
In the CFSS-RDHEI, an encryption key $K_e$ is used to generate the {\color{black}parameters for sharing process, namely the two pseudorandom integer sequences $\bf{Q}$ and $\bf{\tilde{Q}}$ that for image sharing and side information sharing, respectively. Then the security of the orignal image depends on both the number of shares and the encryption key.} Because of the initial state sensitivity, unpredictability and easy implementation, a chaotic system is suitable for generating pseudorandom numbers. Here, an \textit{improved H\`enon Map}\cite{hua2020two} with good performance is used and it is defined as
\begin{equation}
\label{equ.henon}
\begin{cases}
x_{n+1}=(1-\hat{a}x_{n}^{2}+y_{n}) & \mod F ; \\
y_{n+1}=\hat{b}x_{n} & \mod F,
\end{cases}
\end{equation}
where $x_0$ and $y_0$ are two initial values, and $\hat{a}$ and $\hat{b}$ are two control parameters. The encryption key $K_e$ is to initialize $(\hat{a}, \hat{b}, x_0, y_0)$ and the procedure is shown in Algorithm.~\ref{algo.init}.
\begin{algorithm}
	\caption{Initial state generation from the encryption key.}
	\label{algo.init}
	\begin{algorithmic}
		\REQUIRE$F, K_e=[k_1,k_2,\cdots,k_{256}] (k_i  \in \{0,1\})$
		\FOR{$j = 1$ to $4$}
			\STATE$v_j \leftarrow \sum_{i=48j-47}^{48j}k_i\times 2^{-i+48j-48}$
		\ENDFOR
        \FOR{$k = 1$ to $4$}
		\STATE$u_k \leftarrow  \sum_{i=16k-15+192}^{16k+192}k_i\times 2^{i-16k+15-192}$
		\ENDFOR
		\STATE$\hat{a} \leftarrow ((v_1\times u_1) \mod 96) + 5$
		\STATE$\hat{b} \leftarrow ((v_2\times u_2) \mod 96) + 5$
		\STATE$x_0 \leftarrow (v_3\times u_3) \mod F$
		\STATE$y_0 \leftarrow (v_4\times u_4) \mod F$
		\ENSURE Initial state $(\hat{a}, \hat{b}, x_0, y_0)$
	\end{algorithmic}
\end{algorithm}
Then the element of $\bf{Q}$ and $\bf{\tilde{Q}}$ can be obtained as
{\color{black}
\begin{equation}
\label{equ.seq}
\begin{cases}
 {q}_i = (\lfloor x_{i+1} \times 2^{21}\rfloor \mod F-n)+1; \\
 \tilde{q}_i = (\lfloor y_{i+1} \times 2^{21}\rfloor \mod 127-n)+1.
\end{cases}
\end{equation}}

To demonstrate the key sensitivity of the CFSS-RDHEI, an encryption key $K_e^0$ is randomly generated. Then change each bit of the $K_e^0$ to obtain 256 different keys, in which each one has only one bit difference with $K_e^0$. The number of bit change rate (NBCR)~\cite{castro2005strict} is used to calculate the difference of two images. For two \textcolor{black}{encrypted images} $E_1$ and $E_2$ encrypted by two keys with one bit difference, their NBCR is defined as

\begin{equation}
\label{eqa.nbcr}
NBCR = \frac{Ham(E_1, E_2)}{Len},
\end{equation}
where $Ham(E_1, E_2)$ means the Hamming distance of $E_1$ and $E_2$, and $Len$ is the bit length of an image. \textcolor{black}{For two random and independent images $C_1$ and $C_2$, both $C_1(i)$ and $C_2(i)$ have 1/2 probability to be 0 or 1. Then the probabilities of bit pair $(C_1 (i),C_2 (j))$ to be 00, 01, 10, 11 are 1/4, 1/4, 1/4, and 1/4, respectively. Thus, $Ham(C_1 (i),C_2 (i))=1/4+1/4=50\%$. Since this is workable for each bit, the NBCR of two random and independent images is 50\%}.

\begin{figure}[htbp]
	\centering
\begin{minipage}[b]{0.99\linewidth}
	\begin{minipage}[b]{0.49\linewidth}
		\centerline{\includegraphics[width=1\linewidth]{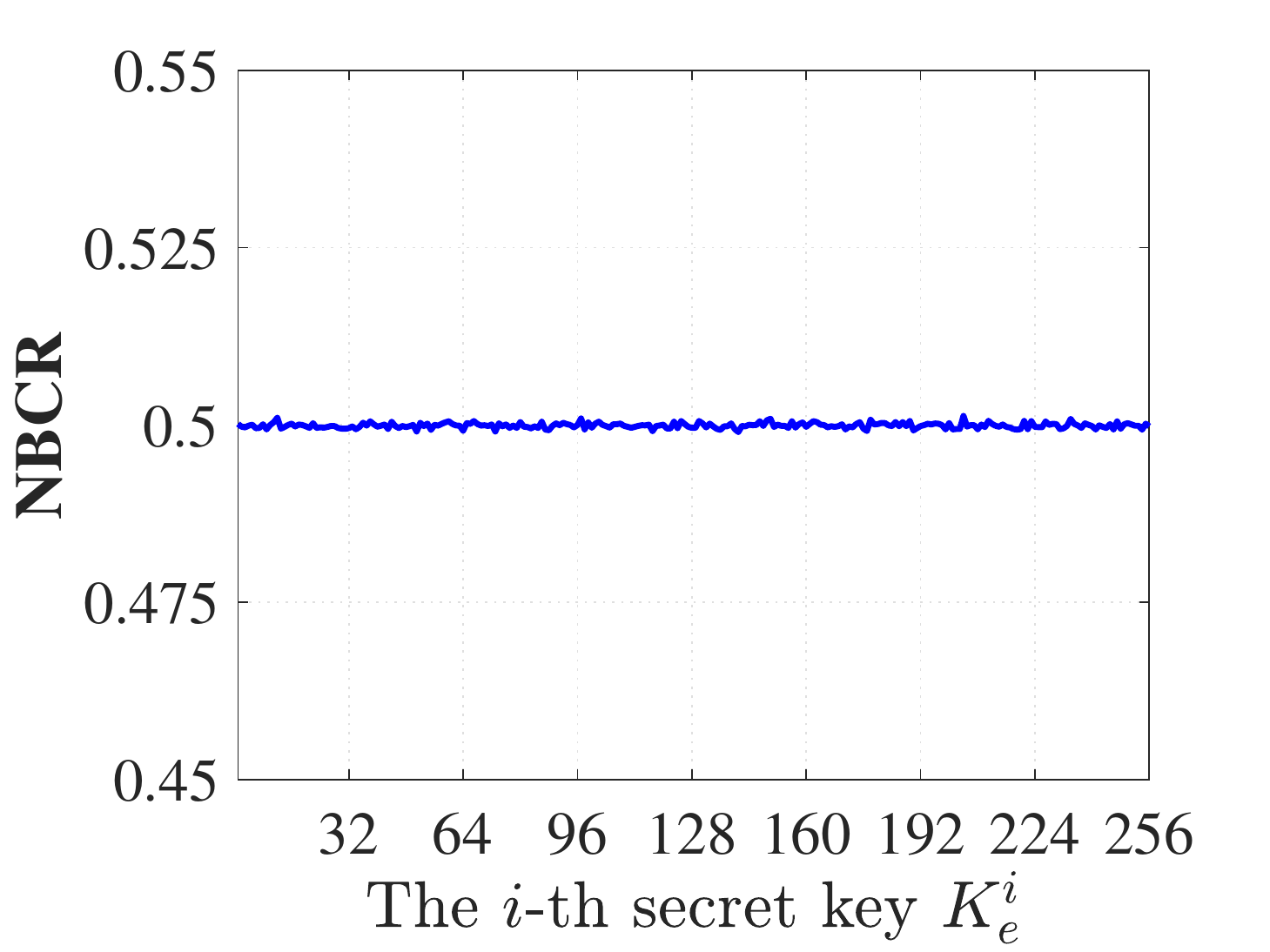}}
		\centerline{(a)}
	\end{minipage}\hfill
	\begin{minipage}[b]{0.49\linewidth}
		\centerline{\includegraphics[width=1\linewidth]{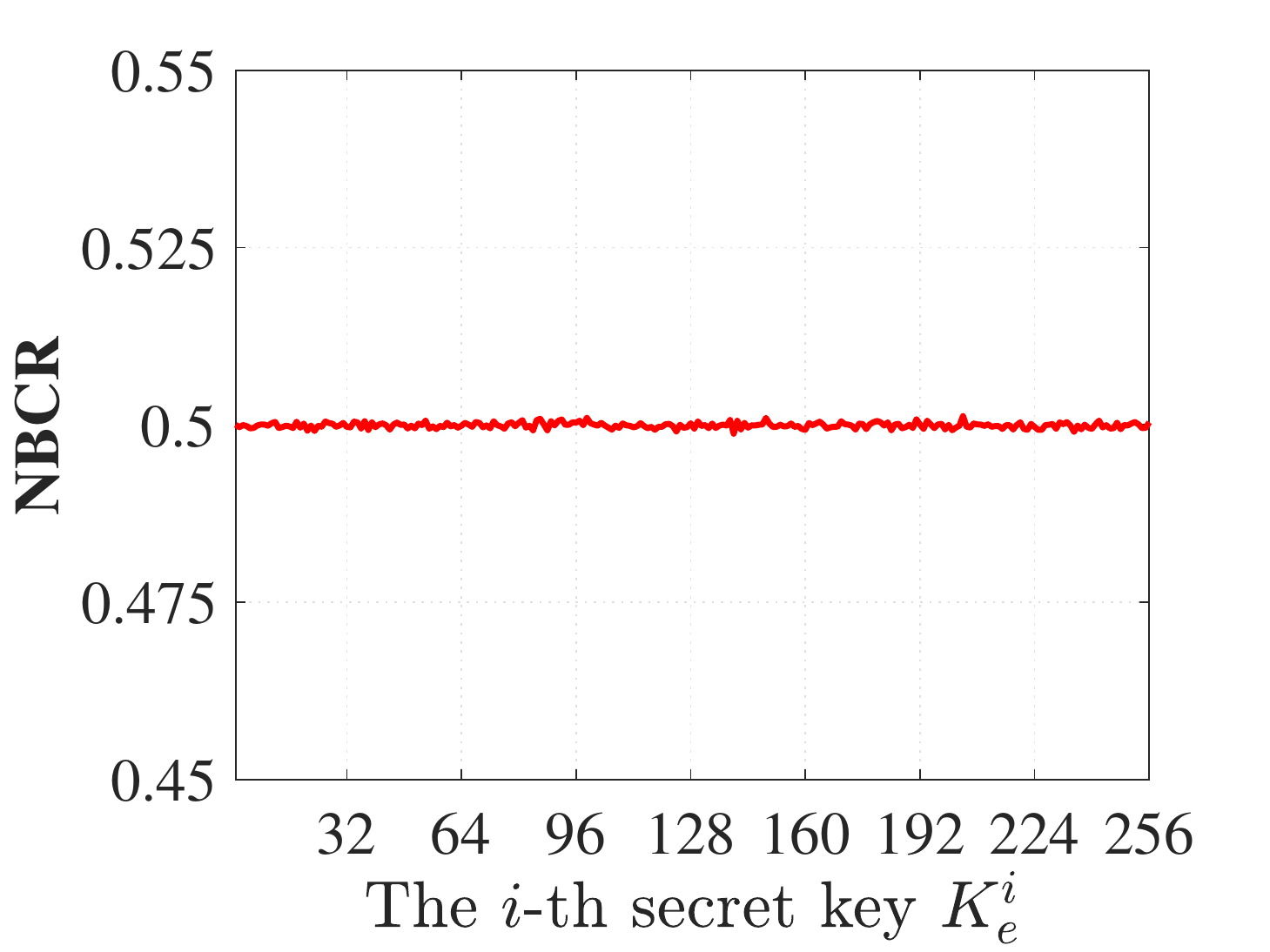}}
		\centerline{(b)}
	\end{minipage}
\end{minipage}\hfill
	\caption{Key sensitivity analysis of the CFSS-RHDEI in each bit of the encryption key. (a) NBCR
 of \textcolor{black}{the first encrypted image}; (b) NBCR of \textcolor{black}{the second encrypted image}.}
	\label{fig.key1}
\end{figure}

Fig.~\ref{fig.key1} shows the experimental results of the key sensitivity analysis in a $(2,2)$-threshold sharing scheme. Each encryption generates two \textcolor{black}{encrypted images} and we calculate the NBCRs of two \textcolor{black}{encrypted images} from two encryption processes with one-bit different keys. This means that the $E_1$ and $E_2$ are two \textcolor{black}{encrypted images} encrypted using one-bit different keys. The results indicate that any one bit change in the encryption key causes the completely different \textcolor{black}{encrypted images}. Therefore, the CFSS-RDHEI is extremely sensitive to its encryption key.

\subsection{Shannon Entropy}
An encrypted image is expected to have uniform-distributed pixels to defense the statistics-based security attacks. The Shannon entropy can be used to measure the distribution of image pixels and it is mathematically defined as
\begin{equation}
  H = -\sum_{i=1}^N Pr(i)\log Pr(i),
\end{equation}
where $N$ is the number of possible pixel values and $Pr(x_i)$ is the probability of the $i$-th possible value. For an 8-bit greyscale image, $N=256$ and the theoretically maximum Shannon entropy can be achieved when each possible value has the same probability, namely $Pr(x_i)=1/256$ for $i\in [1,N]$. Thus, the theoretically maximum Shannon entropy $H_{\max}=-\sum_{i=1}^{256} 1/256\times \log (1/256)=8$. A larger Shannon entropy indicates more uniform distribution of the image pixels.

Table~\ref{table.entropy} lists the Shannon entropies of \textcolor{black}{encrypted images} encrypted by different secret sharing-based RDH-EI methods. The Chen $et~al.$~\cite{chen2019newsecret} method can only generate one \textcolor{black}{encrypted image} for different threshold schemes. For the proposed CFSS-RDHEI, we calculate the Shannon entropies of \textcolor{black}{encrypted images} when the optimal level $l\in \{4,5,6\}$. As can be seen, the \textcolor{black}{CFSS-RDHEI can generate encrypted images that} have larger Shannon entropies than the other three methods and very close to the theoretically maximum value 8. This indicates that \textcolor{black}{its generated encrypted images have} high randomness and cannot display any useful information from pixel distributions.
\begin{table*}[!htbp]
\small
	\renewcommand{\arraystretch}{1}
	\setlength{\tabcolsep}{6pt}
	\begin{center}
		\caption{The Shannon entropies of \textcolor{black}{encrypted images} encrypted by different secret sharing-based methods for the image $Lena$.}
		\label{table.entropy}
		\begin{tabular}{c | c c c| c c c c c c}
			\hline
			\multirow{2}{*}{Methods} & \multicolumn{3}{c|}{(3,3)-threshold scheme}  & \multicolumn{6}{c}{(5,6)-threshold scheme} \\
			\cline{2-4}\cline{5-10}
			& Share 1 & Share 2 & Share 3 & Share 1 & Share 2 & Share 3 & Share 4 & Share 5 & Share 6 \\\hline
			Chen $et~al.$~\cite{chen2019newsecret} & & 7.9708 &  & & & \multicolumn{2}{c}{7.9708} & & \\
			Wu $et~al.$~\cite{wu2018adopting} & 7.9707 & 7.9707 & 7.9709 & 7.9708 & 7.9708 & 7.9708 & 7.9708 & 7.9707 & 7.9707 \\
			Chen $et~al.$~\cite{chen2020secret} & 7.9708 & 7.9709 & 7.9709 & 7.9708 & 7.9708 & 7.9708 & 7.9709 & 7.9708 & 7.9708   \\
			\textbf{CFSS-RDHEI} ($l=4$)  & 7.9972 & 7.9967 & 7.9967 & 7.9955 & 7.9956 & 7.9950 & 7.9956 & 7.9951 & 7.9958 \\
			\textbf{CFSS-RDHEI} ($l=5$)  & 7.9929 & 7.9930 & 7.9961 & 7.9916 & 7.9921 & 7.9912 & 7.9908 & 7.9918 & 7.9917 \\
			\textbf{CFSS-RDHEI} ($l=6$)  & 7.9982 & 7.9981 & 7.9982 & 7.9965 & 7.9968 & 7.9965 & 7.9966 & 7.9965 & 7.9966 \\
			\hline
		\end{tabular}
	\end{center}
\end{table*}
\begin{table*}[!htbp]
\small
	\renewcommand{\arraystretch}{1.1}
	\setlength{\tabcolsep}{2pt}
	\begin{center}
		\caption{NPCR, UACI, SRCC, KRCC Results of CFSS-RDHEI and related methods at (5,6)-threshold scheme.}
		\label{table.diff2}
		\begin{tabular}{c | >{\color{black}}c >{\color{black}}c >{\color{black}}c >{\color{black}}c >{\color{black}}c >{\color{black}}c | >{\color{black}}c >{\color{black}}c >{\color{black}}c >{\color{black}}c >{\color{black}}c >{\color{black}}c}
			\hline
			 & & & \multicolumn{2}{c}{NPCR (\%)} & & & & &  \multicolumn{2}{c}{UACI (\%)} & &   \\\hline
			Chen $et~al.$~\cite{chen2019newsecret} & & & \multicolumn{2}{c}{{\color{black}0.0008}} & & & & &  \multicolumn{2}{c}{{\color{black}0.0005}} & & \\
			Wu $et~al.$~\cite{wu2018adopting} & 0.0008 & 0.0008 & 0.0008 & 0.0008 & 0.0008 & 0.0008 & 0.0003 & 0.0001 & 0.0002 & 0.0002 & 0.0004 & 0.0001 \\
			Chen $et~al.$~\cite{chen2020secret} & 0.0003 & 0.0003 & 0.0003 & 0.0003 & 0.0003 & 0.0003 & 0.0001 & 0.0001 & 0.0002 & 0.0000 & 0.0001 & 0.0000 \\
			\textbf{CFSS-RDHEI} ($l=4$) & 99.6274 & 99.6123 & 99.6785 & 99.6239 & 99.6343 & 995985 & 33.4264 & 33.5621 & 33.6142 & 33.6058 & 33.6721 & 33.5948 \\
			\textbf{CFSS-RDHEI} ($l=5$) & 99.5873 & 99.6306 & 99.5839 & 99.5663 & 99.6274 & 99.5957 & 33.4778 & 33.3290 & 33.3626 & 33.5272 & 33.4318 & 33.3787 \\
			\textbf{CFSS-RDHEI} ($l=6$) & 99.6341 & 99.5869 & 99.6637 & 99.6594 & 99.6334 & 99.5749 & 33.4483 & 33.4045 & 33.3606 & 33.3328 & 33.3973 & 33.4784 \\\hline
			 & & & \multicolumn{2}{c}{SRCC (\%)} & & & & &  \multicolumn{2}{c}{KRCC (\%)} & &   \\\hline
			Chen $et~al.$~\cite{chen2019newsecret} & & & \multicolumn{2}{c}{{\color{black}99.9991}} & & & & &  \multicolumn{2}{c}{{\color{black}99.9999}} & & \\
			Wu $et~al.$~\cite{wu2018adopting} & 99.9993 & 99.9993 & 99.9993 & 99.9993 & 99.9993 & 99.9993 & 99.9999 & 99.9999 & 99.9999 & 99.9999 & 99.9999 & 99.9999 \\
			Chen $et~al.$~\cite{chen2020secret} & 99.9999 & 99.9999 & 99.9999 & 99.9999 & 99.9999 & 99.9999 & 99.9999 & 99.9999 & 99.9999 & 99.9999 & 99.9999 & 99.9999 \\
			\textbf{CFSS-RDHEI} ($l=4$) & -0.0892 & 0.4466 & 0.7054 & 0.7499 & 0.6294 & 0.0326 & -0.0583 & 0.2991 & 0.4726 & 0.5019 & 0.4215 & 0.0224 \\
			\textbf{CFSS-RDHEI} ($l=5$) & 0.3127 & 0.5512 & 0.3707 & -0.2935 & 0.3309 & 0.2614 & 0.0205 & 0.3618 & 0.2462 & -0.1964 & 0.2276 & 0.1798 \\
			\textbf{CFSS-RDHEI} ($l=6$) & -0.1301 & 0.1624 & 0.3375 & 0.6348 & 0.5170 & 0.0039 & -0.0884 & 0.1093 & 0.2258 & 0.4251 & 0.3460 & 0.0025 \\\hline			
		\end{tabular}
	\end{center}
\end{table*}

\subsection{Differential Attack}

The differential attack is an effective cryptanalysis method. It studies how the difference in the plaintext can affect the difference in the ciphertext. By choosing the plaintext to encrypt, the attackers can build the connections between the plaintext and ciphertext, and then use these built connections to reconstruct the ciphertext without secret key.

To resist the differential attack, the slight change in the plaintext should cause large difference in the ciphertext. The number of pixel change rate (NPCR) and uniform average change intensity (UACI)~\cite{wu2011npcruaci} are two measurements to test the ability of a cryptosystem to resist the differential attack. Suppose $E_1$ and $E_2$ of size $M\times N$ are two encrypted images encrypted from two one-bit different original images using the same encryption key. Their NPCR and UACI are defined as
\begin{equation}
\label{eqa.npcr}
NPCR = \frac{\sum_{i=1}^{M}\sum_{j=1}^{N}D(i,j)}{M\times N} \times 100\%
\end{equation}
and
\begin{equation}
\label{eqa.uaci}
UACI= \frac{1}{M\times N}\sum_{i=1}^{M}\sum_{j=1}^{N}\frac{\left | E_1(i,j)-E_2(i,j) \right |}{R} \times 100\%,
\end{equation}
respectively, where $R$ is the largest possible value (255 for the 8-bit greyscale image) and $D(i,j)$ is the difference of the two images and defined as
\begin{equation}
\label{eqa.npcr2}
D(i,j)=\begin{cases}
1 & \quad \mbox{for}\ E_1(i,j) \neq E_2(i,j);\\
0 & \quad \mbox{otherwise}.
\end{cases}
\end{equation}

As can be seen, the NPCR is the total number of different pixels in two images and the UACI means the average difference between pixels in two images. \textcolor{black}{According to the theoretical analysis in~\cite{wu2011npcruaci}, the ideal NPCR and UACI values for 8-bit greyscale image are $99.609\%$ and $33.464\%$, respectively}.

To investigate the correlation of the \textcolor{black}{encrypted images} $E_1$ and $E_2$, we use the Spearman's rank correlation coefficient (SRCC) and Kendall rank correlation coefficient (KRCC)~\cite{wang2010srcckrcc} to evaluate their monotonicity. The two correlation coefficients are defined as
\begin{equation}
\label{eqa.srcc}
SRCC = 1-\frac{6\sum_{i=1}^{M\times N}d_i^2}{(M\times N)\times ((M\times N)^2-1)}
\end{equation}
and
\begin{equation}
\label{eqa.krcc}
KRCC=\frac{N_c - N_d}{(M\times N)\times ((M\times N)-1)/2},
\end{equation}
respectively, where $d_i$ is the difference between ranks of the two \textcolor{black}{encrypted images}, and $N_c$ and $N_d$ are the numbers of concordant and discordant pairs in two images.

The obtained SRCC and KRCC are within the range $[-1, 1]$. The value that closes to -1 or 1 means a high correlation between two images and a values close to 0 indicates a low correlation.

Table~\ref{table.diff2} lists the results in $(5,6)$-threshold. It can be seen from the tables that the other secret sharing-based RDH-EI methods obtain '0' in NPCR and UACI tests while '1' in SRCC and KRCC tests. This indicates that the slight change in the original image cannot cause significant difference in the encrypted results. Then these \textcolor{black}{traditional secret sharing-based} encryption strategies cannot resist the differential attack. On the other hand, the proposed CFSS-RDHEI can obtain NPCR and UACI that close to the theoretical values, namely $33.464\%$ and $99.609\%$, and the SRCC and KRCC values are all close to 0. \textcolor{black}{Because the CFSS uses a randomly selected previous sharing result to share the current section and any random integer can be used to share the first section. In each execution, the selections of previous sharing result and random integer are different. Thus, even though the modified pixel is the last one of the image, the obtained encrypted results are also different. Thus, the CFSS-RDHEI has high ability to resist the differential attack.}

\subsection{Missing Share Attack}
Our CFSS-RHDEI uses the CFSS to encrypt the original image to be $n$ \textcolor{black}{encrypted images}. Only using $r$ $(r\leq n)$ \textcolor{black}{encrypted images}, one can completely recover the original image. This property is guaranteed by Eq.~\eqref{eqa.share} that at least $r$ equations are required to solve the $r$ coefficients $a_0,...,a_{r-1}$ of the polynomial. Assuming that an attacker has collected $r-1$ \textcolor{black}{encrypted images}, $r-1$ equations can be constructed. The probability of solving $r$ variables using $r-1$ equations is $1/F$. For an image of size $M\times N$, there are $(M\times N)/(r-1)$ sections. Then, the probability of correctly recover the original image using $r-1$ \textcolor{black}{encrypted images} is $(1/F)^{M\times N /(r-1)}$. Thus, the attackers are difficult to correctly recover the original image using $r-1$ \textcolor{black}{encrypted images}.

\begin{figure}[!htbp]
	\centering
	\begin{minipage}[b]{1.0\linewidth}
		\begin{minipage}[b]{0.25\linewidth}
			\centerline{\includegraphics[width=1\linewidth]{boat.png}}
			\centerline{(a)}
		\end{minipage}\hfill
		\begin{minipage}[b]{0.25\linewidth}
			\centerline{\includegraphics[width=0.5\linewidth]{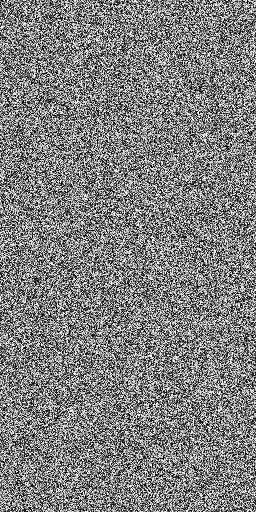}}
			\centerline{(b)}
		\end{minipage}\hfill
		\begin{minipage}[b]{0.25\linewidth}
			\centerline{\includegraphics[width=0.5\linewidth]{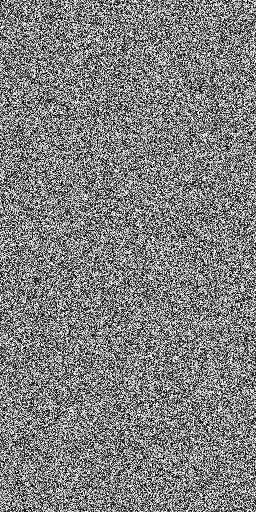}}
		    \centerline{(c)}
		\end{minipage}\hfill
		\begin{minipage}[b]{0.25\linewidth}
			\centerline{\includegraphics[width=0.5\linewidth]{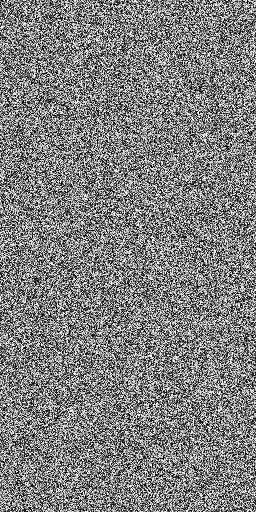}}
			\centerline{(d)}
		\end{minipage}
		\begin{minipage}[b]{0.25\linewidth}
			\centerline{\includegraphics[width=0.5\linewidth]{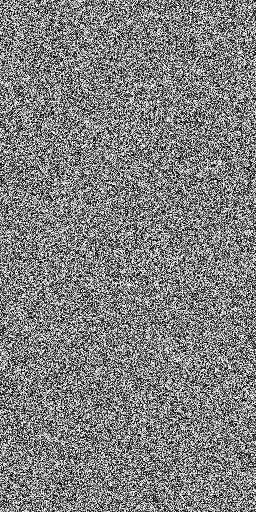}}
			\centerline{(e)}
		\end{minipage}\hspace{10pt}
		\begin{minipage}[b]{0.25\linewidth}
			\centerline{\includegraphics[width=1\linewidth]{boat.png}}
			\centerline{(f)}
		\end{minipage}\hspace{20pt}
		\begin{minipage}[b]{0.25\linewidth}
			\centerline{\includegraphics[width=1\linewidth]{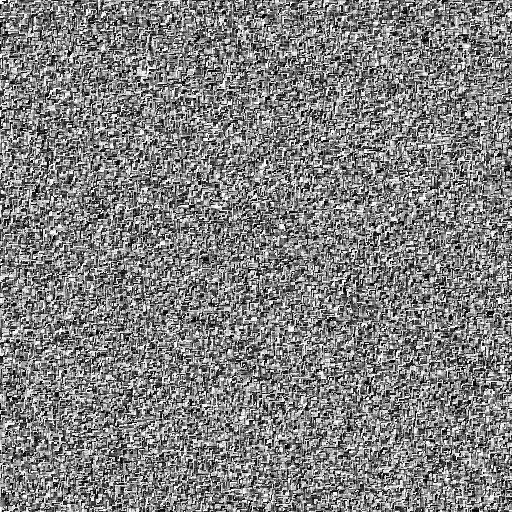}}
			\centerline{(g)}
		\end{minipage}
	\end{minipage}
	\caption{The construction of original image using fake \textcolor{black}{encrypted image} when $r=3$. (a) The original image of size $512\times 512$; (b)-(d) three \textcolor{black}{encrypted images} of size $512\times 256$ generated by CFSS-RDHEI; (e) a randomly generated fake \textcolor{black}{encrypted image}; (f) the reconstructed image using (b), (c), and (d); (g) the reconstructed image using (b), (c) and (e).}
	\label{fig.fakeshare}
\end{figure}

Assuming that the attacker would like to reconstruct the original image using $r-1$ true \textcolor{black}{encrypted images}. Since $r$ equations are required to solve the $r$ coefficients of the polynomial, a fake \textcolor{black}{encrypted image} is constructed with the same size of the true \textcolor{black}{encrypted images}. Fig.~\ref{fig.fakeshare} shows the experimental results of reconstructing the original image using $r-1$ true \textcolor{black}{encrypted images} and one fake \textcolor{black}{encrypted image}. As can be seen, only using the $r$ true \textcolor{black}{encrypted images}, the original images can be correctly reconstructed. Using one fake \textcolor{black}{encrypted image}, one cannot obtain any information about the original image. {\color{black}All the secret sharing-based RDH-EI schemes can defend against the missing share attack.}

\section{Conclusion}
\label{Section6}

In this paper, we first developed a new polynomial-based secret sharing scheme called CFSS. It adopts the cipher-feedback strategy to process the image pixel and thus strictly follows the cryptography standards, which results in a high security level. Using the developed CFSS, we further proposed a high-security RDH-EI scheme with multiple data hiders called CFSS-RDHEI. A multi-MSBs prediction method without {\color{black}insecure} error flag is used to embed secret data. First, an optimal level for each image is set using its prediction {\color{black}precision}. Then, the side information is generated to store the required information of recovering the multi-MSBs of the image. Next, the side information and multi-LSBs of the image are encrypted to several shares by CFSS, and each share of the side information is embedded into one \textcolor{black}{encrypted image}. Finally, the content owner send each obtained \textcolor{black}{encrypted image} with side information embedded to one data hider, which can embed secret data into the \textcolor{black}{encrypted image}. When collecting a predefined number of \textcolor{black}{encrypted images}, one can completely recover the original image. Performance emulations showed that the proposed CFSS-RDHEI has high capacity rate and its generated \textcolor{black}{encrypted images} are much smaller than other secret sharing-based RDH-EI schemes. Security analysis demonstrates that it has a high security level to defense some commonly used security attacks and can outperform other similar schemes.

%

\end{document}